\documentclass[english,pra,aps,superscriptaddress,longbibliography,preprint]{revtex4-1}
\usepackage[T1]{fontenc}
\usepackage[latin9]{inputenc}
\usepackage{mathrsfs}
\usepackage{mathtools}
\usepackage{amsmath}
\usepackage{amssymb}
\usepackage{graphicx}

\makeatletter
\usepackage[colorlinks,citecolor=blue,linkcolor=blue]{hyperref}

\makeatother

\usepackage{babel}
\begin{document}
\title{Quantum-classical correspondence principle for heat distribution in
quantum Brownian motion}
\author{Jin-Fu Chen}
\affiliation{School of Physics, Peking University, Beijing, 100871, China}
\author{Tian Qiu}
\affiliation{School of Physics, Peking University, Beijing, 100871, China}
\author{H. T. Quan}
\thanks{Corresponding author: htquan@pku.edu.cn}
\affiliation{School of Physics, Peking University, Beijing, 100871, China}
\affiliation{Collaborative Innovation Center of Quantum Matter, Beijing 100871,
China}
\affiliation{Frontiers Science Center for Nano-optoelectronics, Peking University,
Beijing, 100871, China}
\date{\today}
\begin{abstract}
Quantum Brownian motion, described by the Caldeira-Leggett model,
brings insights to understand phenomena and essence of quantum thermodynamics,
especially the quantum work and heat associated with their classical
counterparts. By employing the phase-space formulation approach, we
study the heat distribution of a relaxation process in the quantum
Brownian motion model. The analytical result of the characteristic
function of heat is obtained at any relaxation time with an arbitrary
friction coefficient. By taking the classical limit, such a result
approaches the heat distribution of the classical Brownian motion
described by the Langevin equation, indicating the quantum-classical
correspondence principle for heat distribution. We also demonstrate
that the fluctuating heat at any relaxation time satisfies the exchange
fluctuation theorem of heat, and its long-time limit reflects complete
thermalization of the system. Our research brings justification for
the definition of the quantum fluctuating heat via two-point measurements.
\end{abstract}
\maketitle

\section{Introduction}

In the past few decades, the discovery of fluctuation theorems \citep{Gallavotti1995a,Jarzynski1997,Crooks1999,Jarzynski2004}
and the establishment of the framework of stochastic thermodynamics
\citep{Jarzynski2011,Sekimoto2010,Seifert2012} deepened our understanding
about the fluctuating nature of thermodynamic quantities (such as
work, heat and entropy production) in microscopic systems \citep{Seifert2005,Esposito2009,Campisi2011,Klages2013,Horodecki2013,Ciliberto2017}.
Among various fluctuation theorems, the non-equilibrium work relation
\citep{Jarzynski1997} sharpens our understanding of the second law
of thermodynamics by presenting an elegant and precise equality associating
the free energy change with the fluctuating work. Such a relation
was later extended to the quantum realm based on the two-point measurement
definition of the quantum fluctuating work \citep{Tasaki2000,Kurchan2000},
soon after its discovery in the classical regime. The work statistics
has been widely studied in various microscopic classical and quantum
systems \citep{Talkner2007,Deffner2008,Liu2014,Zhu2016,Funo2018a,Salazar2019a,Jarzynski2015,Deffner2016,GarciaMata2017,Fei2020,Qiu2020}.
Historically, the quantum-classical correspondence principle played
an essential role in the development of the theory of quantum mechanics
and the interpretation of the transition from quantum to classical
world \citep{Zurek1991,Zurek2003}. In Refs. \citep{Jarzynski2015,Zhu2016,GarciaMata2017},
it is demonstrated that the existence of the quantum-classical correspondence
principle for work distribution brings justification for the definition
of quantum fluctuating work via two-point measurements.

Compared to work statistics, heat statistics relevant to thermal transport
associated with a nonequilibrium stationary state has been extensively
studied \citep{Saito2007,Dubi2011,Thingna2012,Wang2013,Thingna2016,He2016,Segal2016,Kilgour2019,Wang2017,Aurell2020},
but the heat statistics in a finite-time quantum thermodynamic process
\citep{Denzler2018,Salazar2019,Popovic2021} and its quantum-classical
correspondence have been less explored. A challenge is that the precise
description of the bath dynamics requires handling a huge number of
degrees of freedom of the heat bath. Different approaches have been
proposed to calculate the quantum fluctuating heat and its statistics,
such as the non-equilibrium Green's function approach to quantum thermal
transport \citep{Saito2007,KarstenBalzer2012,Wang2013,Esposito2015a,Kilgour2019,Polanco2021}
and the path-integral approach to quantum thermodynamics \citep{Aron2010,Mallick2011,Carrega2015,Funo2018,Yeo2019}.
However, very few analytical results about the heat statistics have
been obtained for the relaxation processes in open quantum systems.
These analytical results are limited to either the relaxation dynamics
described by the Lindblad master equation \citep{Denzler2018,Salazar2019}
or the long-time limit independent of the relaxation dynamics \citep{Fogedby2020}.
On the other hand, some results about the heat statistics in the classical
Brownian motion model have been reported \citep{Zon2004,Imparato2007,Fogedby2009,Chatterjee2010,GomezSolano2011,Salazar2016,Pagare2019,Paraguassu2021a,Gupta2021}.
How the quantum and the classical heat statistics (especially associated
with the relaxation dynamics in finite time) are related to each other
has not been explored so far, probably due to the difficulty in studying
the heat statistics in open quantum systems \citep{Esposito2015,Talkner2016a,Talkner2020}.

In this article, we study the heat statistics of a quantum Brownian
motion model described by the Caldeira-Leggett Hamiltonian \citep{Bez1980,Caldeira1983,Caldeira1983b,Unruh1989,Breuer2007,Weiss2008,Funo2018},
where the heat bath is modeled as a collection of harmonic oscillators.
Although it is well known that the dynamics of such an open quantum
system can approach that of the classical Brownian motion in the classical
limit $\hbar\rightarrow0$ \citep{Caldeira1983}, less is known about
the heat statistics of this model during the finite-time relaxation
process. We here focus on the relaxation process without external
driving (the Hamiltonian of the system is time-independent), and the
quantum fluctuating heat can thus be defined as the difference of
the system energy between the initial and the final measurements \footnote{Usually the quantum fluctuating heat is defined via two-point measurements
over the heat bath. When the Hamiltonian of the system is time-independent,
the internal energy change of the system is completely caused by the
heat exchange. The quantum fluctuating heat can thus be alternatively
defined via two-point measurements over the system, whose number of
degrees of freedom is much smaller than that of the heat bath. Hence,
the calculation of the heat statistics can be significantly simplified
under this definition.}. Under the Ohmic spectral density, the dynamics of the composite
system is exactly solvable in the continuum limit of the bath oscillators
\citep{Yu1994}. By employing the phase-space formulation approach
\citep{Wigner1932,Hillery_1984,Polkovnikov2010}, we obtain analytical
results of the characteristic function of heat for the Caldeira-Leggett
model at any relaxation time $\tau$ with an arbitrary friction coefficient
$\kappa$. Previously, such an approach was employed to study the
quantum corrections to work \citep{Fei2018,Qian2019,Brodier2020}
and entropy \citep{Qiu2020a,Qiu2021}. Analytical results of the heat
statistics bring important insights to understand the fluctuating
property of heat. By taking the classical limit $\hbar\rightarrow0$,
the heat statistics of the Caldeira-Leggett model approaches that
of the classical Brownian motion model. Thus, our results verify the
quantum-classical correspondence principle for heat distribution,
and provide justification for the definition of the quantum fluctuating
heat via two-point measurements. We also verify from the analytical
results that the heat statistics satisfies the exchange fluctuation
theorem of heat \citep{Jarzynski2004}.

The rest of this article is organized as follows. In Sec. \ref{sec:The-characteristic-function},
we introduce the Caldeira-Leggett model and define the quantum fluctuating
heat. In Sec. \ref{sec:The-results-of}, the analytical results of
the characteristic function of heat are obtained by employing the
phase-space formulation approach. We show the quantum-classical correspondence
of the heat distribution, and discuss the heat distribution in the
long-time limit or with the extremely weak or strong coupling strength.
The conclusion is given in Sec. \ref{sec:Conclusion}.

\section{the Caldeira-Leggett model and the heat statistics\label{sec:The-characteristic-function}}

\subsection{The Caldeira-Leggett model}

The quantum Brownian motion is generally described by the Caldeira-Leggett
model \citep{Caldeira1983,Caldeira1983b}, where the system is modeled
as a single particle moving in a specific potential, and the heat
bath is a collection of harmonic oscillators. For simplicity, we choose
the harmonic potential for the system \citep{Unruh1989,Hu1992,Karrlein1997,Ford2001},
where the dynamics of such an open quantum system can be solved analytically.
The system will relax to the equilibrium state at the temperature
of the heat bath. We study the heat distribution of such a quantum
relaxation process, and obtain analytically the characteristic function
of heat and its classical correspondence based on the phase-space
formulation of quantum mechanics.

The total Hamiltonian of the composite system is $H_{\mathrm{tot}}=H_{S}+H_{B}+H_{SB}$
with each term

\begin{align}
H_{S} & =\frac{1}{2}\frac{\hat{p}_{0}^{2}}{m_{0}}+\frac{1}{2}m_{0}\omega_{0}^{2}\hat{q}_{0}^{2}\\
H_{B} & =\sum_{n=1}^{N}\left(\frac{1}{2}\frac{\hat{p}_{n}^{2}}{m_{n}}+\frac{1}{2}m_{n}\omega_{n}^{2}\hat{q}_{n}^{2}\right)\\
H_{SB} & =-\hat{q}_{0}\sum_{n=1}^{N}\left(C_{n}\hat{q}_{n}\right)+\sum_{n=1}^{N}\left(\frac{C_{n}^{2}}{2m_{n}\omega_{n}^{2}}\hat{q}_{0}^{2}\right),\label{eq:Hsb}
\end{align}
where $m_{0}$, $\omega_{0}$, $\hat{q}_{0}$, $\hat{p}_{0}$ ($m_{n}$,
$\omega_{n}$, $\hat{q}_{n}$, $\hat{p}_{n}$ with $n=1,2,3,...,N$)
are the mass, frequency, position and momentum of the system (the
$n$-th bath harmonic oscillator), and $C_{n}$ is the coupling strength
between the system and the $n$-th bath harmonic oscillator. The counter-term
$\sum_{n}[C_{n}^{2}/(2m_{n}\omega_{n}^{2})]\hat{q}_{0}^{2}$ is included
in the interaction Hamiltonian $H_{SB}$ to cancel the frequency shift
of the system.

The spectral density is defined as $J(\omega)\coloneqq\sum_{n}[C_{n}^{2}/(2m_{n}\omega_{n})]\delta(\omega-\omega_{n})$.
We adopt an Ohmic spectral density with the Lorentz-Drude cutoff \citep{Breuer2007}
\begin{equation}
J(\omega)=\frac{m_{0}\kappa}{\pi}\omega\frac{\Omega_{0}^{2}}{\Omega_{0}^{2}+\omega^{2}},\label{eq:135-1}
\end{equation}
where $\kappa$ is the friction coefficient. A sufficiently large
cutoff frequency $\Omega_{0}$ ($\Omega_{0}\gg\omega_{0}$) is applied
to ensure a finite counter-term, and the dynamics with the timescale
exceeding $1/\Omega_{0}$ is Markovian. Under such a spectral density,
the dissipation dynamics of the Caldeira-Leggett model with a weak
coupling strength $\kappa\ll\omega_{0}$ reproduces that of the classical
underdamped Brownian motion when taking the classical limit $\hbar\rightarrow0$
\citep{Caldeira1983}.

We assume the initial state to be a product state of the system and
the heat bath
\begin{equation}
\rho(0)=\rho_{S}(0)\otimes\rho_{B}^{G},\label{eq:initial_product_state}
\end{equation}
which makes it possible to define the quantum fluctuating heat via
two-point measurements. Here, $\rho_{S}(0)$ is the initial state
of the system, and $\rho_{B}^{G}=\exp(-\beta H_{B})/Z_{B}(\beta)$
is the Gibbs distribution of the heat bath with the inverse temperature
$\beta$ and the partition function $Z_{B}(\beta)={\rm Tr}[\exp(-\beta H_{B})]$.

\subsection{The quantum fluctuating heat in the relaxation process}

We study the heat distribution of the relaxation process based on
the two-point measurement definition of the quantum fluctuating heat.
When no external driving is applied to the system, the Hamiltonian
of the system is time-independent. Since no work is performed during
the relaxation process, the quantum fluctuating heat can be defined
as

\begin{equation}
Q_{l^{\prime}l}=E_{l^{\prime}}^{S}-E_{l}^{S},\label{eq:Qnn}
\end{equation}
where $E_{l}^{S}$ ($E_{l^{\prime}}^{S}$) is the eigenenergy of the
system corresponding to the outcome $l$ ($l^{\prime}$) at the initial
(final) time $t=0$ ($t=\tau$). The two-point measurements over the
heat bath can be hardly realized due to a huge number of degrees of
freedom of the heat bath \citep{Funo2018a}, while the measurements
over the small quantum system are much easier in principle. The positive
sign corresponds to the energy flowing from the heat bath to the system.

For the system prepared in an equilibrium state, no coherence exists
in the initial state, and the initial density matrix of the system
commutes with the Hamiltonian of the system, $[\rho(0),H_{S}]=0$.
The probability of observing the transition from $l$ and $l^{\prime}$
is

\begin{equation}
p_{\tau,l^{\prime}l}=\gamma_{\tau,l^{\prime}l}p_{l},
\end{equation}
with the conditional transition probability $\gamma_{\tau,l^{\prime}l}={\rm Tr}\left[(\hat{P}_{l^{\prime}}^{S}\otimes I_{B})U_{\mathrm{tot}}(\tau)(\hat{P}_{l}^{S}\otimes\rho_{B}^{G})U_{\mathrm{tot}}^{\dagger}(\tau)\right]$
and the initial probability $p_{l}=\mathrm{Tr}[\rho(0)\hat{P}_{l}^{S}]$.
Here, $\hat{P}_{l}^{S}=\left|l\right\rangle \left\langle l\right|$
is the projection operator corresponding to the outcome $l$. The
heat distribution is defined as

\begin{equation}
P_{\tau}(q)\coloneqq\sum_{l^{\prime},l}\delta(q-Q_{l^{\prime}l})p_{\tau,l^{\prime}l}.\label{eq:PtauQ}
\end{equation}
The characteristic function of heat $\chi_{\tau}(\nu)$ is defined
as the Fourier transform of the heat distribution $\chi_{\tau}(\nu)\coloneqq\sum_{l^{\prime},l}\exp[i\nu(E_{l^{\prime}}^{S}-E_{l}^{S})]p_{\tau,l^{\prime}l}$,
which can be rewritten explicitly as

\begin{equation}
\chi{}_{\tau}(\nu)={\rm Tr}\left[e^{i\nu H_{S}}U_{\mathrm{tot}}(\tau)\left(e^{-i\nu H_{S}}\rho(0)\right)U_{\mathrm{tot}}^{\dagger}(\tau)\right],\label{eq:chiQ_definition}
\end{equation}
where $U_{\mathrm{tot}}(\tau)=\exp(-iH_{\mathrm{tot}}\tau/\hbar)$
is the unitary time-evolution operator of the composite system.

Our goal is to analytically calculate the characteristic function
$\chi{}_{\tau}(\nu)$. Previously, the quantum-classical correspondence
principle for heat statistics has been analyzed with the path-integral
approach to quantum thermodynamics \citep{Funo2018}, yet the explicit
result of the characteristic function (or generating function) of
heat has not been obtained so far. We employ the phase-space formulation
approach to solve this problem, and rewrite the characteristic function
Eq. (\ref{eq:chiQ_definition}) into 
\begin{equation}
\chi{}_{\tau}(\nu)={\rm Tr}\left[e^{i\nu H_{S}^{\mathrm{H}}(\tau)}\eta(0)\right],\label{eq:chiQ_in_phasespaceformulation}
\end{equation}
where the system Hamiltonian in the Heisenberg picture is 
\begin{equation}
H_{S}^{\mathrm{H}}(\tau)=U_{\mathrm{tot}}^{\dagger}(\tau)H_{S}U_{\mathrm{tot}}(\tau),
\end{equation}
and the density-matrix-like operator $\eta(0)$ is

\begin{equation}
\eta(0)=\left[e^{-i\nu H_{S}}\rho_{S}(0)\right]\otimes\rho_{B}^{G}.
\end{equation}

We express Eq. (\ref{eq:chiQ_in_phasespaceformulation}) with the
phase-space formulation of quantum mechanics \citep{Wigner1932,Hillery_1984,Polkovnikov2010,Fei2018,Qian2019,Brodier2020}

\begin{equation}
\chi_{\tau}(\nu)=\frac{1}{(2\pi\hbar)^{N+1}}\int{\rm d}\mathbf{z}\left[e^{i\nu H_{S}^{\mathrm{H}}(\tau)}\right]_{w}(\mathbf{z})\cdot P(\mathbf{z}),\label{eq:chiQ}
\end{equation}
where $\mathbf{z}$ represents a point $\mathbf{z}=[\mathbf{q},\mathbf{p}]=[q_{0},...,q_{N},p_{0},...,p_{N}]$
in the phase space of the composite system, and the integral is performed
over the whole phase space. The subscript ``$w$'' indicates the
Weyl symbol of the corresponding operator, and $P(\mathbf{z})$ is
the Weyl symbol of the operator $\eta(0)$, which is explicitly defined
as \citep{Wigner1932} 
\begin{equation}
P(\mathbf{z}):=\int d\mathbf{y}\left\langle \mathbf{q}-\frac{\mathbf{y}}{2}\Big|\eta(0)\Big|\mathbf{q}+\frac{\mathbf{y}}{2}\right\rangle e^{\frac{i\mathbf{p}\cdot\mathbf{y}}{\hbar}}.\label{eq:108a}
\end{equation}
In the following, we will calculate the heat statistics Eq. (\ref{eq:chiQ})
by employing the phase-space formulation approach.

\section{Results of the characteristic function of heat \label{sec:The-results-of}}

We show a sketch of the derivation of the heat statistics $\chi_{\tau}(\nu)$
with the details left in Appendix \ref{Appendix:derivations}. We
specifically consider the system is initially prepared at an equilibrium
state $\rho_{S}(0)=\exp(-\beta^{\prime}H_{S})/Z_{S}(\beta^{\prime})$
with the inverse temperature $\beta^{\prime}$ and the partition function
$Z_{S}(\beta^{\prime})=1/[2\sinh(\beta^{\prime}\hbar\omega_{0}/2)]$.
The heat bath is at the inverse temperature $\beta$, which is different
from $\beta^{\prime}$. In Eq. (\ref{eq:chiQ}), the two Weyl symbols
$\left[e^{i\nu H_{S}^{\mathrm{H}}(\tau)}\right]_{w}(\mathbf{z})$
and $P(\mathbf{z})$ are obtained as

\begin{equation}
\left[e^{i\nu H_{S}^{\mathrm{H}}(\tau)}\right]_{w}(\mathbf{z})=\frac{1}{\cos\left(\frac{\nu\hbar\omega_{0}}{2}\right)}\exp\left[\frac{i}{2\hbar}\mathbf{z}^{\mathrm{T}}\boldsymbol{\tilde{\Lambda}}_{\nu z}(\tau)\mathbf{z}\right],\label{eq:einuHsH_wignerfunction}
\end{equation}
and

\begin{equation}
P(\mathbf{z})=\frac{2\sinh\left(\frac{\beta^{\prime}\hbar\omega_{0}}{2}\right)}{\cosh\left[\frac{(\beta^{\prime}+i\nu)\hbar\omega_{0}}{2}\right]}\cdot\left[\prod_{n=1}^{N}2\tanh\left(\frac{\beta\hbar\omega_{n}}{2}\right)\right]\cdot\exp\left(-\frac{1}{2\hbar}\mathbf{z}^{\mathrm{T}}\boldsymbol{\Lambda}_{\beta z}\mathbf{z}\right),\label{eq:Pz_wignerfunction}
\end{equation}
where the explicit expressions of the matrices $\boldsymbol{\tilde{\Lambda}}_{\nu z}(\tau)$
and $\boldsymbol{\Lambda}_{\beta z}$ are given in Eqs. (\ref{eq:33})
and (\ref{eq:Lambda_=00005Cbetaz}), respectively.

Substituting Eqs. (\ref{eq:einuHsH_wignerfunction}) and (\ref{eq:Pz_wignerfunction})
into Eq. (\ref{eq:chiQ}), the characteristic function of heat at
any relaxation time $\tau$ with an arbitrary friction coefficient
$\kappa$ is finally obtained as

\begin{align}
\chi_{\tau}(\nu)= & \left\{ \left[(1+i\Xi)(1-i\Theta\Xi)-i\Xi(1-\Theta-i\Theta\Xi)\frac{\kappa^{2}\cos\left(2\hat{\omega}_{0}\tau\right)-4\omega_{0}^{2}}{(\kappa^{2}-4\omega_{0}^{2})e^{\kappa\tau}}\right]^{2}\right.\nonumber \\
 & \left.+\Xi^{2}(1-\Theta-i\Theta\Xi)^{2}\left[\left(\frac{\kappa^{2}\cos\left(2\hat{\omega}_{0}\tau\right)-4\omega_{0}^{2}}{(\kappa^{2}-4\omega_{0}^{2})e^{\kappa\tau}}\right)^{2}-e^{-2\kappa\tau}\right]\right\} ^{\frac{1}{2}},\label{eq:characteristic_function_CL_model}
\end{align}
where the quantities $\Xi$ and $\Theta$ are

\begin{align}
\Xi & =\frac{\tan\left(\frac{\nu\hbar\omega_{0}}{2}\right)}{\tanh\left[\frac{(\beta^{\prime}+i\nu)\hbar\omega_{0}}{2}\right]-i\tan\left(\frac{\nu\hbar\omega_{0}}{2}\right)},\label{eq:Upsilon}\\
\Theta & =\frac{\tanh\left[\frac{(\beta^{\prime}+i\nu)\hbar\omega_{0}}{2}\right]-i\tan\left(\frac{\nu\hbar\omega_{0}}{2}\right)}{\tanh\left(\frac{\beta\hbar\omega_{0}}{2}\right)}.\label{eq:Theta}
\end{align}
Induced by the friction, the frequency of the system harmonic oscillator
is shifted to $\hat{\omega}_{0}=\sqrt{\omega_{0}^{2}-\kappa^{2}/4}$.

From the analytical results of the heat statistics Eq. (\ref{eq:characteristic_function_CL_model}),
the average heat $\left\langle Q\right\rangle (\tau)=\left.-i\partial_{\nu}[\ln\chi_{\tau}(\nu)]\right|_{\nu=0}$
is immediately obtained as

\begin{align}
\left\langle Q\right\rangle (\tau) & =\frac{\omega_{0}\hbar}{2}\left[\coth\left(\frac{\beta\omega_{0}\hbar}{2}\right)-\coth\left(\frac{\beta^{\prime}\omega_{0}\hbar}{2}\right)\right]\left[1-\frac{\kappa^{2}\cos\left(2\hat{\omega}_{0}\tau\right)-4\omega_{0}^{2}}{(\kappa^{2}-4\omega_{0}^{2})e^{\kappa\tau}}\right],
\end{align}
and the variance $\mathrm{Var}\left(Q\right)(\tau)=\left.-\partial_{\nu}^{2}[\ln\chi_{\tau}(\nu)]\right|_{\nu=0}$
is

\begin{equation}
\mathrm{Var}\left(Q\right)(\tau)=\mathrm{I}+\mathrm{II}\cdot e^{-\kappa\tau}+\mathrm{III}\cdot e^{-2\kappa\tau},
\end{equation}
with

\begin{align}
\mathrm{I} & =\frac{\omega_{0}^{2}\hbar^{2}\left[\text{csch}^{2}\left(\frac{\beta\omega_{0}\hbar}{2}\right)+\text{csch}^{2}\left(\frac{\beta^{\prime}\omega_{0}\hbar}{2}\right)\right]}{4},\\
\mathrm{II} & =\frac{\kappa^{2}\cos(2\hat{\omega}_{0}\tau)-4\omega_{0}^{2}}{2\hat{\omega}_{0}^{2}}\cdot\frac{\omega_{0}^{2}\hbar^{2}\left[\coth^{2}\left(\frac{\beta\omega_{0}\hbar}{2}\right)+\text{csch}^{2}\left(\frac{\beta^{\prime}\omega_{0}\hbar}{2}\right)-\coth\left(\frac{\beta\omega_{0}\hbar}{2}\right)\coth\left(\frac{\beta^{\prime}\omega_{0}\hbar}{2}\right)\right]}{4}\\
\mathrm{III} & =\frac{\kappa^{4}\cos(4\hat{\omega}_{0}\tau)+8\omega_{0}^{2}\kappa^{2}[1-2\cos(2\hat{\omega}_{0}\tau)]+16\omega_{0}^{4}}{16\hat{\omega}_{0}^{4}}\cdot\frac{\omega_{0}^{2}\hbar^{2}\left[\coth\left(\frac{\beta\omega_{0}\hbar}{2}\right)-\coth\left(\frac{\beta^{\prime}\omega_{0}\hbar}{2}\right)\right]^{2}}{4}.
\end{align}
Similarly, one can calculate the higher cumulants from the analytical
results of the heat statistics. In the following, we will examine
the properties of the heat statistics of the quantum Brownian motion.

\subsection{Quantum-classical correspondence principle for heat statics and the
exchange fluctuation theorem of heat}

We further take the classical limit $\hbar\rightarrow0$, or more
rigorously $\beta\hbar\omega_{0}\rightarrow0$. The two quantities
approach $\Xi\rightarrow\nu/\beta^{\prime}$ and $\Theta\rightarrow\beta^{\prime}/\beta$,
and the characteristic function of heat {[}Eq. (\ref{eq:characteristic_function_CL_model}){]}
becomes
\begin{align}
\chi_{\tau}^{\mathrm{cl}}(\nu)= & \left\{ \left[(1+i\frac{\nu}{\beta^{\prime}})(1-i\frac{\nu}{\beta})-\frac{i\nu\left(\beta-\beta^{\prime}-i\nu\right)}{\beta\beta^{\prime}}\frac{\left(\kappa^{2}\cos\left(2\hat{\omega}_{0}\tau\right)-4\omega_{0}^{2}\right)}{(\kappa^{2}-4\omega_{0}^{2})e^{\kappa\tau}}\right]^{2}\right.\nonumber \\
 & \left.+\nu^{2}\left(\frac{\beta-\beta^{\prime}-i\nu}{\beta\beta^{\prime}}\right)^{2}\left[\left(\frac{\kappa^{2}\cos\left(2\hat{\omega}_{0}\tau\right)-4\omega_{0}^{2}}{(\kappa^{2}-4\omega_{0}^{2})e^{\kappa\tau}}\right)^{2}-e^{-2\kappa\tau}\right]\right\} ^{-\frac{1}{2}},\label{eq:characteristic_function_classical_result}
\end{align}
which is consistent with the results obtained from the classical Brownian
motion described by the Kramers equation (see Ref. \citep{Paraguassu2021a}
or Appendix \ref{sec:characteristic-function-of}). The average heat
is

\begin{align}
\left\langle Q^{\mathrm{cl}}\right\rangle (\tau) & =\frac{\beta^{\prime}-\beta}{\beta\text{\ensuremath{\beta^{\prime}}}}\left[1-\frac{\kappa^{2}\cos\left(2\hat{\omega}_{0}\tau\right)-4\omega_{0}^{2}}{(\kappa^{2}-4\omega_{0}^{2})e^{\kappa\tau}}\right],
\end{align}
and the variance $\mathrm{Var}\left(Q^{\mathrm{cl}}\right)(\tau)=\left.-\partial_{\nu}^{2}[\ln\chi_{\tau}^{\mathrm{cl}}(\nu)]\right|_{\nu=0}$
is

\begin{align}
\mathrm{Var}\left(Q^{\mathrm{cl}}\right)(\tau) & =\mathrm{I}^{\mathrm{cl}}+\mathrm{II}^{\mathrm{cl}}\cdot e^{-\kappa\tau}+\mathrm{III}^{\mathrm{cl}}\cdot e^{-2\kappa\tau},
\end{align}
with

\begin{align}
\mathrm{I}^{\mathrm{cl}} & =\frac{\beta^{2}+\text{\ensuremath{\beta^{\prime2}}}}{\beta^{2}\beta^{\prime2}}\\
\mathrm{II}^{\mathrm{cl}} & =\frac{\kappa^{2}\cos(2\hat{\omega}_{0}\tau)-4\omega_{0}^{2}}{2\hat{\omega}_{0}^{2}}\cdot\frac{\beta^{2}-\beta\beta^{\prime}+\text{\ensuremath{\beta^{\prime2}}}}{\beta^{2}\beta^{\prime2}}\\
\mathrm{III}^{\mathrm{cl}} & =\frac{\kappa^{4}\cos(4\hat{\omega}_{0}\tau)+8\omega_{0}^{2}\kappa^{2}[1-2\cos(2\hat{\omega}_{0}\tau)]+16\omega_{0}^{4}}{16\hat{\omega}_{0}^{4}}\cdot\frac{(\beta-\beta^{\prime})^{2}}{\beta^{2}\beta^{\prime2}}.
\end{align}

From Eq. (\ref{eq:characteristic_function_CL_model}) {[}or the classical
counterpart Eq. (\ref{eq:characteristic_function_classical_result}){]},
one can see the characteristic function of heat exhibits the following
symmetry

\begin{equation}
\chi_{\tau}(\nu)=\chi_{\tau}[-i(\beta-\beta^{\prime})-\nu],
\end{equation}
which shows the heat distribution satisfies the exchange fluctuation
theorem of heat in the differential form $P_{\tau}(Q)/P_{\tau}(-Q)=\exp[-(\beta-\beta^{\prime})Q]$
\citep{Jarzynski2004}. By setting $\nu=0$, we obtain the relation
$\chi_{\tau}[-i(\beta-\beta^{\prime})]=\chi_{\tau}(0)=1$, which is
exactly the exchange fluctuation theorem of heat in the integral form
$\left\langle \exp[(\beta-\beta^{\prime})Q]\right\rangle =1$.

\subsection{Long-time limit}

In the long-time limit $\tau\rightarrow\infty$, the characteristic
functions of heat {[}Eqs. (\ref{eq:characteristic_function_CL_model})
and (\ref{eq:characteristic_function_classical_result}){]} become

\begin{align}
\chi_{\infty}(\nu) & =\frac{\left(1-e^{-\beta^{\prime}\omega_{0}\hbar}\right)\left(1-e^{-\beta\omega_{0}\hbar}\right)}{\left(1-e^{-(\beta^{\prime}+i\nu)\omega_{0}\hbar}\right)\left(1-e^{-(\beta-i\nu)\omega_{0}\hbar}\right)},\label{eq:chiinfquantum}
\end{align}
and

\begin{equation}
\chi_{\infty}^{\mathrm{cl}}(\nu)=\frac{\beta^{\prime}\beta}{(\beta^{\prime}+i\nu)(\beta-i\nu)}.\label{eq:chiinfclassical}
\end{equation}
Such results, independent of the relaxation dynamics, are in the form
\begin{equation}
\chi_{\mathrm{th}}(\nu)=\frac{Z_{S}(\beta^{\prime}+i\nu)Z_{S}(\beta-i\nu)}{Z_{S}(\beta^{\prime})Z_{S}(\beta)},\label{eq:chiQ_thermalization}
\end{equation}
reflecting complete thermalization of the system \citep{Fogedby2009}.
For example, the relaxation of a harmonic oscillator governed by the
quantum-optical master equation gives the identical characteristic
function of heat in the long-time limit \citep{Denzler2018}. In Appendix
\ref{sec:heat_characteristic_function_complete_thermalization}, we
demonstrate that the characteristic function of heat for any relaxation
processes with complete thermalization is always in the form of Eq.
(\ref{eq:chiQ_thermalization}). With the simple expressions (\ref{eq:chiinfquantum})
and (\ref{eq:chiinfclassical}) of the characteristic functions, the
heat distributions are obtained from the inverse Fourier transform
as

\begin{equation}
P_{\infty}(q)=\begin{cases}
\frac{\left(1-e^{-\beta^{\prime}\omega_{0}\hbar}\right)\left(1-e^{-\beta\omega_{0}\hbar}\right)}{1-e^{-(\beta^{\prime}+\beta)\omega_{0}\hbar}}\sum_{j=0}^{\infty}\delta(q-j\omega_{0}\hbar)e^{-\beta q} & q\geq0\\
\frac{\left(1-e^{-\beta^{\prime}\omega_{0}\hbar}\right)\left(1-e^{-\beta\omega_{0}\hbar}\right)}{1-e^{-(\beta^{\prime}+\beta)\omega_{0}\hbar}}\sum_{j=1}^{\infty}\delta(q+j\omega_{0}\hbar)e^{\beta^{\prime}q} & q<0
\end{cases},\label{eq:P_infq_quantum}
\end{equation}
and

\begin{equation}
P_{\infty}^{\mathrm{cl}}(q)=\begin{cases}
\frac{\beta^{\prime}\beta}{\beta^{\prime}+\beta}e^{-\beta q} & q\geq0\\
\frac{\beta^{\prime}\beta}{\beta^{\prime}+\beta}e^{\beta^{\prime}q} & q<0
\end{cases},\label{eq:P_infq_classical}
\end{equation}
which are exactly the same as the long-time results obtained in Ref.
\citep{Denzler2018}.

\subsection{Weak/Strong-coupling limit in finite time}

In the weak-coupling limit $\kappa\ll\omega_{0}$, the characteristic
function of heat {[}Eq. (\ref{eq:characteristic_function_CL_model}){]}
becomes

\begin{align}
\chi_{\tau}^{\mathrm{w}}(\nu) & =\frac{1}{(1+i\Xi)(1-i\Xi\Theta)(1-e^{-\kappa\tau})+e^{-\kappa\tau}}.\label{eq:weak_characteristic_function}
\end{align}
There is only one relaxation timescale associated to $\kappa$. Such
situation corresponds to the highly underdamped regime of the classical
Brownian motion, and a systematic method has been proposed to study
the heat distribution \citep{Salazar2016} as well as the work distribution
under an external driving \citep{Salazar2020,Chen2021a}.

In the strong coupling limit $\kappa\gg\omega_{0}$, the characteristic
function of heat {[}Eq. (\ref{eq:characteristic_function_CL_model}){]}
becomes

\begin{align}
\chi_{\tau}^{\mathrm{s}}(\nu)= & \frac{1}{\sqrt{\left(1+i\Xi\right)\left(1-i\Xi\Theta\right)\left(1-e^{-2\kappa\tau}\right)+e^{-2\kappa\tau}}}\nonumber \\
 & \times\frac{1}{\sqrt{\left(1+i\Xi\right)\left(1-i\Xi\Theta\right)\left(1-e^{-\frac{2\omega_{0}^{2}}{\kappa}\tau}\right)+e^{-\frac{2\omega_{0}^{2}}{\kappa}\tau}}}.\label{eq:strong_charateristic_function}
\end{align}
The relaxation timescales of the momentum (the first factor) and the
coordinate (the second factor) are separated. The long-time limits
of both Eqs. (\ref{eq:weak_characteristic_function}) and (\ref{eq:strong_charateristic_function})
are equal to Eq. (\ref{eq:chiinfquantum}). In classical thermodynamics,
the usual overdamped approximation neglects the motion of the momentum,
hence the heat statistics derived under such an approximation is incomplete
\citep{Imparato2007}. Actually, the momentum degree of freedom also
contributes to the heat statistics.

\subsection{Numerical results\label{subsec:Numerical-results}}

\begin{figure}
\includegraphics[width=16cm]{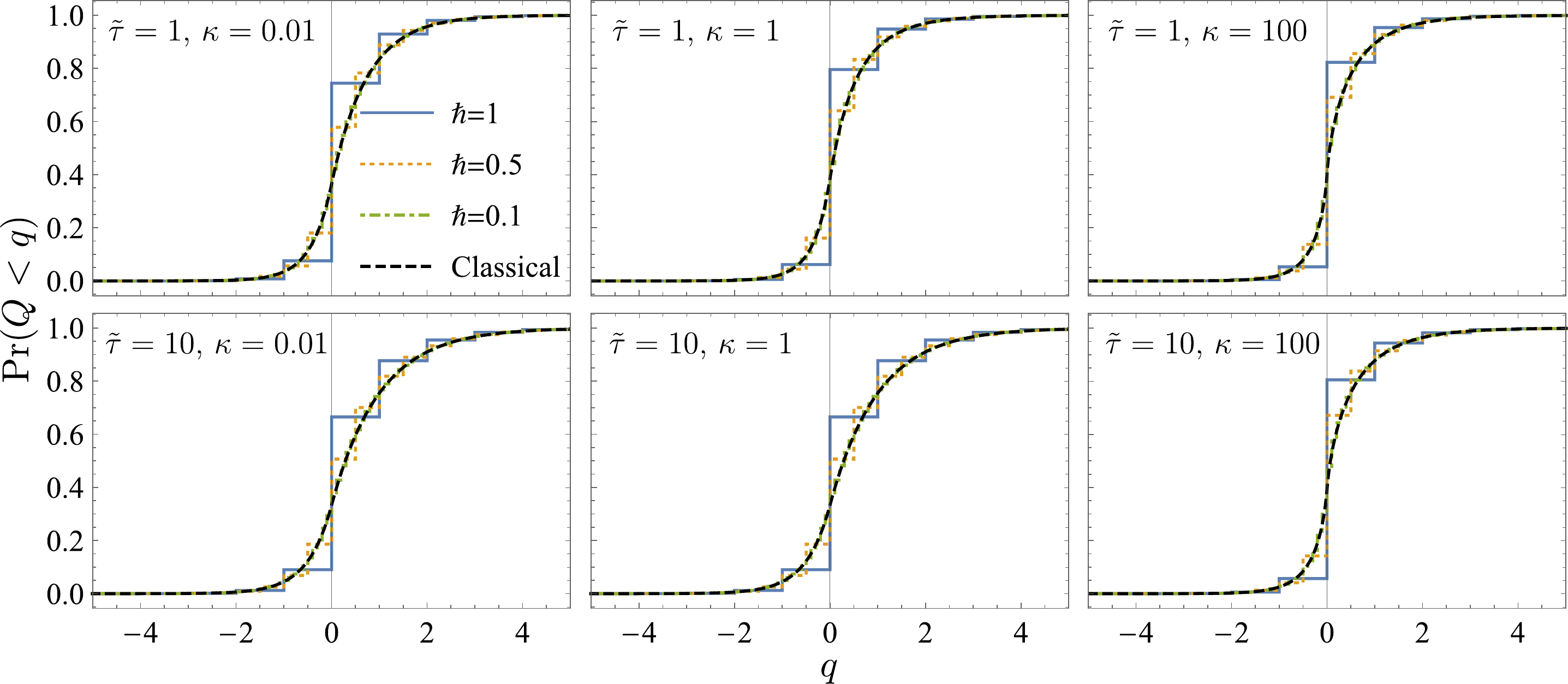}

\caption{The cumulative heat distribution function $\mathrm{Pr}(Q<q)$. The
choices of the parameters are given in the main text. We compare the
results of the Caldeira-Leggett model (blue solid, orange dotted and
green dot-dashed curves) in Eq. (\ref{eq:characteristic_function_CL_model})
and those of the classical Brownian motion (black dashed curve) in
Eq. (\ref{eq:characteristic_function_classical_result}). The rescaled
relaxation time is $\tilde{\tau}=\kappa\tau=1$ in the upper subfigures
and $\tilde{\tau}=10$ in the lower subfigures. The left, middle and
right subfigures illustrate the results for the weak ($\kappa=0.01$),
intermediate ($\kappa=1$) and strong coupling strength ($\kappa=100$).
\label{fig:evolution_characteristic_function-1}}
\end{figure}

In Fig. \ref{fig:evolution_characteristic_function-1}, we show the
cumulative heat distribution function $\mathrm{Pr}(Q<q)\coloneqq\int_{-\infty}^{q}P_{\tau}(q^{\prime})dq^{\prime}$
with different friction coefficients $\kappa=0.01,\:1$ and $100$
at the rescaled relaxation time $\tilde{\tau}=\kappa\tau=1$ and $10$.
We set the mass $m_{0}=1$ and the frequency $\omega_{0}=1$ for the
system harmonic oscillator, the inverse temperatures $\beta=1$ and
$\beta^{\prime}=2$ for the initial equilibrium states of the heat
bath and the system, respectively. The Planck constant is set to be
$\hbar=1,\:0.5,\:0.1$. With the decrease of $\hbar$, the quantum
results {[}Eq. (\ref{eq:characteristic_function_CL_model}){]} approaches
the classical result {[}Eq. (\ref{eq:characteristic_function_classical_result}){]}.
The quantum-classical correspondence of the heat distribution is thus
demonstrated for generic values of the friction coefficient $\kappa$.

For $\kappa=0.01$ and $1$, complete thermalization is achieved at
$\tilde{\tau}=10$. The left-lower and middle-lower subfigures show
the identical distribution characterized by Eqs. (\ref{eq:P_infq_quantum})
and (\ref{eq:P_infq_classical}). For $\kappa=100$, the momentum
degree of freedom has been thermalized {[}$\exp(-2\tilde{\tau})\approx0$
in Eq. (\ref{eq:strong_charateristic_function}){]} while the coordinate
degree of freedom remains frozen {[}$\exp[-2(\omega_{0}^{2}/\kappa^{2})\tilde{\tau}]\approx1$
in Eq. (\ref{eq:strong_charateristic_function}){]}. Thus, the distribution
in the right-lower subfigure is different from the middle-lower subfigure.

\begin{figure}
\includegraphics[width=16cm]{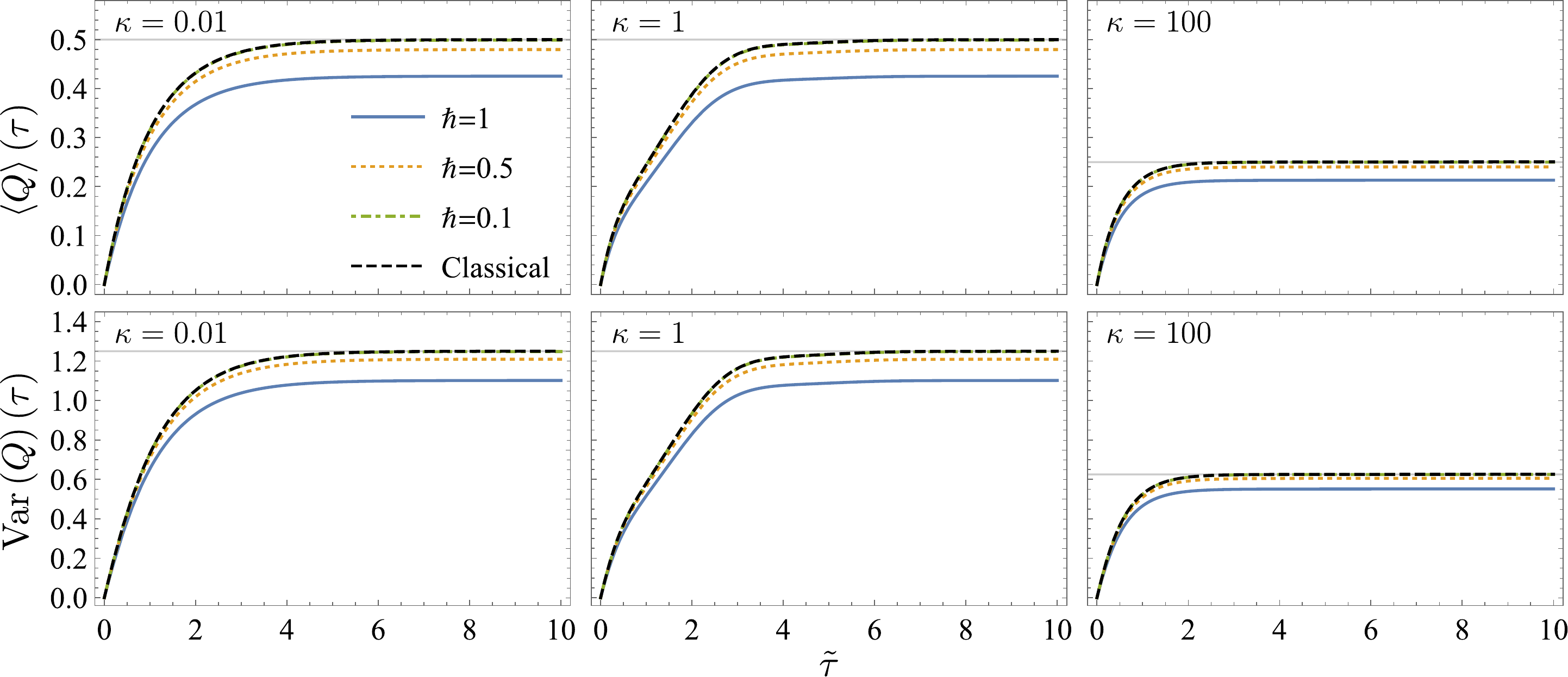}\caption{The evolution of the mean value $\left\langle Q\right\rangle (\tau)$
(upper subfigures) and the variance $\mathrm{Var}\left(Q\right)(\tau)$
(lower subfigures) of the heat statistics as functions of the rescaled
time $\tilde{\tau}=\kappa\tau$.\label{fig:The-evolution-ofQand_varainceQ}}
\end{figure}

In Fig. \ref{fig:The-evolution-ofQand_varainceQ}, we illustrate the
results of the mean value $\left\langle Q\right\rangle (\tau)$ and
the variance $\mathrm{Var}\left(Q\right)(\tau)$ with different friction
coefficients $\kappa=0.01,\:1$ and $100$. The parameters are the
same as those in Fig. \ref{fig:evolution_characteristic_function-1}.
The quantum results approach the classical results\textbf{ }with the
decrease of $\hbar$. For $\kappa=0.01$ and $1$ (left and middle
subfigures), complete thermalization is reached when $\tilde{\tau}>5$.
The mean value and the variance approach $\underset{\tau\rightarrow\infty}{\lim}\left\langle Q^{\mathrm{cl}}\right\rangle (\tau)=1/\beta-1/\beta^{\prime}$
and $\underset{\tau\rightarrow\infty}{\lim}\mathrm{Var}\left(Q^{\mathrm{cl}}\right)(\tau)=1/\beta^{2}+1/\beta^{\prime2}$(gray
horizontal lines). For $\kappa=100$ (right subfigures), only the
momentum degree of freedom is thermalized at this timescale. Thus,
the mean value and the variance take half value of their long-time
limits. When the coordinate degree of freedom is also thermalized
in the long-time limit ($\tilde{\tau}\gg\kappa^{2}/\omega_{0}^{2}=10^{4}$),
the mean value and the variance are expected to approach the same
values as those in the middle subfigures.

\section{Conclusion\label{sec:Conclusion}}

Previously, the heat statistics of the relaxation processes has been
studied analytically in open quantum systems described by the Lindblad
master equation \citep{Denzler2018,Salazar2019,Fogedby2020}. However,
due to the rotating wave approximation and other approximations. Such
quantum systems do not possess a well-defined classical counterpart.
Hence, the quantum-classical correspondence principle for heat distribution
has not been well established.

In this paper, we study the heat statistics of the quantum Brownian
motion model described by the Caldeira-Leggett Hamiltonian, in which
the bath dynamics is explicitly considered. By employing the phase-space
formulation approach, we obtain the analytical expressions of the
characteristic function of heat at any relaxation time $\tau$ with
an arbitrary friction coefficient $\kappa$. Analytical results of
heat statistics bring important insights to the studies of quantum
thermodynamics. For example, in the classical limit, our results approach
the heat statistics of the classical Brownian motion. Thus, the quantum-classical
correspondence principle for heat statistics is verified in this model.
Our analytical results provide justification for the definition of
quantum fluctuating heat via two-point measurements.

We have also discussed the characteristic function of heat in the
long-time limit or with the extremely weak/strong coupling strength.
In the long-time limit, the form of the characteristic function of
heat reflects complete thermalization of the system. In addition,
from the analytical expressions of the heat statistics, we can immediately
verify the exchange fluctuation theorem of heat. The phase-space formulation
can be further utilized to study the joint statistics of work and
heat in a driven open quantum system, which will be beneficial to
explore the fluctuations of power and efficiency in finite-time quantum
heat engines.
\begin{acknowledgments}
H. T. Quan acknowledges support from the National Natural Science
Foundation of China under Grants No. 11775001, No. 11534002, and No.
11825001. This paper is dedicated to Prof. Wojciech Zurek on the occasion
of his 70th birthday for his kind and generous supports to one of
the authors (H. T. Quan), and for his many insightful contributions
to our understanding about the quantum to classical transition.
\end{acknowledgments}

\appendix

\section{Derivation to the characteristic function of heat (\ref{eq:characteristic_function_CL_model})
\label{Appendix:derivations}}

We show the detailed derivation to the characteristic function of
heat $\chi_{\tau}(\nu)$. We first calculate the two Wigner functions
$\left[e^{i\nu H_{S}^{\mathrm{H}}(t)}\right]_{w}(\mathbf{z})$ and
$P(\mathbf{z})$. Then the final result Eq. (\ref{eq:characteristic_function_CL_model})
is obtained from Eq. (\ref{eq:chiQ}).

\subsection{$\left[e^{i\nu H_{S}^{\mathrm{H}}(t)}\right]_{w}(\mathbf{z})$}

With the quadratic Hamiltonian $H_{S}^{\mathrm{H}}(t)$, the Wigner
function $\left[e^{i\nu H_{S}^{\mathrm{H}}(t)}\right]_{w}(\mathbf{z})$
is \citep{Ford2001,Qiu2021}

\begin{align}
\left[e^{i\nu H_{S}^{\mathrm{H}}(t)}\right]_{w}(\mathbf{z}) & =\frac{1}{\cos\left(\frac{\omega_{0}\hbar\nu}{2}\right)}\exp\left[i\frac{m_{0}\omega_{0}}{\hbar}\tan\left(\frac{\omega_{0}\hbar\nu}{2}\right)q_{0}^{2}(t)+i\frac{1}{m_{0}\hbar\omega_{0}}\tan\left(\frac{\omega_{0}\hbar\nu}{2}\right)p_{0}^{2}(t)\right]\label{eq:14}\\
 & =\frac{1}{\cos\left(\frac{\omega_{0}\hbar\nu}{2}\right)}\exp\left[\frac{i}{2\hbar}\mathbf{z}^{\mathrm{T}}(t)\boldsymbol{\Lambda}_{\nu z}\mathbf{z}(t)\right],
\end{align}
where $\mathbf{z}(t)$ gives the trajectory in the phase space determined
by the initial point $\mathbf{z}(0)=\mathbf{z}$, and $\boldsymbol{\Lambda}_{\nu z}$
is a rank-2 diagonal matrix

\begin{equation}
\boldsymbol{\Lambda}_{\nu z}=\left(\begin{array}{cccc}
2m_{0}\omega_{0}\tan\left(\frac{\omega_{0}\hbar\nu}{2}\right)\\
 & \mathbf{O}\\
 &  & \frac{2}{m_{0}\omega_{0}}\tan\left(\frac{\omega_{0}\hbar\nu}{2}\right)\\
 &  &  & \mathbf{O}
\end{array}\right),
\end{equation}
with an $N\times N$ zero matrix $\mathbf{O}$. The unlisted elements
are zeros. The trajectory $\mathbf{z}(t)$ satisfies the classical
equation of motion (also the equation of motion in the Heisenberg
picture)

\begin{align}
\dot{q}_{0} & =\frac{p_{0}}{m_{0}},\label{eq:35}\\
\dot{q}_{n} & =\frac{p_{n}}{m_{n}},\\
\dot{p}_{0} & =-m_{0}\tilde{\omega}_{0}^{2}q_{0}+\sum_{n}C_{n}q_{n},\\
\dot{p}_{n} & =-m_{n}\omega_{n}^{2}q_{n}+C_{n}q_{0},\label{eq:38}
\end{align}
with $\tilde{\omega}_{0}^{2}=\omega_{0}^{2}+\sum_{n=1}^{N}C_{n}^{2}/(m_{0}m_{n}\omega_{n}^{2}).$
The above differential equations can be rewritten into a compact form
$\dot{\mathbf{z}}(t)=\mathbf{L}\mathbf{z}(t)$. The trajectory $\mathbf{z}(t)=[\mathbf{q}(t),\mathbf{p}(t)]$
in the phase space characterizes the evolution of the composite system
with the positions $\mathbf{q}(t)=[q_{0}(t),...,q_{N}(t)]$ and the
momenta $\mathbf{p}(t)=[p_{0}(t),...,p_{N}(t)]$, and is related to
the initial point by the dynamical map $\mathbf{z}(t)=\exp(\mathbf{L}t)\mathbf{z}(0).$
The $(2N+2)\times(2N+2)$ matrix $\mathbf{L}$ is explicitly 
\begin{equation}
\mathbf{L}=\left(\begin{array}{cccccccccc}
 &  &  &  &  & \frac{1}{m_{0}}\\
 &  &  &  &  &  & \frac{1}{m_{1}}\\
 &  &  &  &  &  &  & \frac{1}{m_{2}}\\
 &  &  &  &  &  &  &  & ...\\
 &  &  &  &  &  &  &  &  & \frac{1}{m_{N}}\\
-m_{0}\tilde{\omega}_{0}^{2} & C_{1} & C_{2} & ... & C_{N}\\
C_{1} & -m_{1}\omega_{1}^{2}\\
C_{2} &  & -m_{2}\omega_{2}^{2}\\
... &  &  & ...\\
C_{N} &  &  &  & -m_{N}\omega_{N}^{2}
\end{array}\right),\label{eq:Lmat}
\end{equation}
and the matrix exponential is formally written as

\begin{equation}
\exp\left(\mathbf{L}t\right)=\left(\begin{array}{cccccccc}
\alpha_{0} & \alpha_{1} & ... & \alpha_{N} & \frac{\beta_{0}}{m_{0}} & \frac{\beta_{1}}{m_{1}} & ... & \frac{\beta_{N}}{m_{N}}\\
\gamma_{1} & \Lambda_{11} & ... & \Lambda_{1N} & \frac{\xi_{1}}{m_{0}} & \frac{\Delta_{11}}{m_{1}} & ... & \frac{\Delta_{1N}}{m_{N}}\\
... & ... & ... & ... & ... & ... & ... & ...\\
\gamma_{N} & \Lambda_{N1} & ... & \Lambda_{NN} & \frac{\xi_{N}}{m_{0}} & \frac{\Delta_{N1}}{m_{1}} & ... & \frac{\Delta_{NN}}{m_{N}}\\
m_{0}\dot{\alpha}_{0} & m_{0}\dot{\alpha}_{1} & ... & m_{0}\dot{\alpha}_{N} & \frac{m_{0}}{m_{0}}\dot{\beta}_{0} & \frac{m_{0}}{m_{1}}\dot{\beta}_{1} & ... & \frac{m_{0}}{m_{N}}\dot{\beta}_{N}\\
m_{1}\dot{\gamma}_{1} & m_{1}\dot{\Lambda}_{11} & ... & m_{1}\dot{\Lambda}_{1N} & \frac{m_{1}}{m_{0}}\dot{\xi}_{1} & \frac{m_{1}}{m_{1}}\dot{\Delta}_{11} & ... & \frac{m_{1}}{m_{N}}\dot{\Delta}_{1N}\\
... & ... & ... & ... & ... & ... & ... & ...\\
m_{N}\dot{\gamma}_{N} & m_{N}\dot{\Lambda}_{N1} & ... & m_{N}\dot{\Lambda}_{NN} & \frac{m_{N}}{m_{0}}\dot{\xi}_{N} & \frac{m_{N}}{m_{1}}\dot{\Delta}_{N1} & ... & \frac{m_{N}}{m_{N}}\dot{\Delta}_{NN}
\end{array}\right).\label{eq:expLt}
\end{equation}
We rewrite the quadratic form into $\mathbf{z}^{\mathrm{T}}(t)\boldsymbol{\Lambda}_{\nu z}\mathbf{z}(t)=\mathbf{z}^{\mathrm{T}}(0)\boldsymbol{\tilde{\Lambda}}_{\nu z}(t)\mathbf{z}(0)$
with 
\begin{align}
\boldsymbol{\tilde{\Lambda}}_{\nu z}(t) & =\exp\left(\mathbf{L}^{\mathrm{T}}t\right)\boldsymbol{\Lambda}_{\nu z}\exp\left(\mathbf{L}t\right).\label{eq:33}
\end{align}

We next carry out every element in Eq. (\ref{eq:expLt}) through the
Laplace transforms of Eqs. (\ref{eq:35})-(\ref{eq:38})

\begin{align}
s\tilde{q}_{0}(s)-q_{0}(0) & =\frac{\tilde{p}_{0}(s)}{m_{0}},\\
s\tilde{q}_{n}(s)-q_{n}(0) & =\frac{\tilde{p}_{n}(s)}{m_{n}},\\
s\tilde{p}_{0}(s)-p_{0}(0) & =-m_{0}\tilde{\omega}_{0}^{2}\tilde{q}_{0}(s)+\sum_{n}C_{n}\tilde{q}_{n}(s),\\
s\tilde{p}_{n}(s)-p_{n}(0) & =-m_{n}\omega_{n}^{2}\tilde{q}_{n}(s)+C_{n}\tilde{q}_{0}(s).
\end{align}
Representing $\tilde{q}_{n}(s)$ and $\tilde{p}_{n}(s)$ with $\tilde{q}_{0}(s)$
and the initial conditions, we obtain

\begin{equation}
\left\{ s^{2}+\tilde{\omega}_{0}^{2}-\sum_{n}\left[\frac{C_{n}^{2}}{m_{0}m_{n}\left(s^{2}+\omega_{n}^{2}\right)}\right]\right\} \tilde{q}_{0}(s)=\dot{q}_{0}(0)+sq_{0}(0)+\sum_{n}\frac{C_{n}}{m_{0}}\left[\frac{\dot{q}_{n}(0)+sq_{n}(0)}{s^{2}+\omega_{n}^{2}}\right].\label{eq:68}
\end{equation}
Under the Ohmic spectral density {[}Eq. (\ref{eq:135-1}){]}, the
above equation is simplified to

\begin{equation}
(s^{2}+\kappa s+\omega_{0}^{2})\tilde{q}_{0}(s)=\dot{q}_{0}(0)+sq_{0}(0)+\sum_{n}\frac{C_{n}}{m_{0}}\left[\frac{\dot{q}_{n}(0)+sq_{n}(0)}{s^{2}+\omega_{n}^{2}}\right],\label{eq:s^2+=00005Cetas}
\end{equation}
where the summation on the left-hand side of Eq. (\ref{eq:68}) can
be approximately expressed as

\begin{align}
\sum_{n}\left[\frac{C_{n}^{2}}{m_{0}m_{n}\left(s^{2}+\omega_{n}^{2}\right)}\right] & \approx-\kappa s+\sum_{n}\frac{C_{n}^{2}}{m_{0}m_{n}\omega_{n}^{2}},
\end{align}
with a large cutoff frequency $\Omega_{0}$. The inverse Laplace transform
gives the differential equation of $q_{0}(t)$ as

\begin{equation}
\ddot{q}_{0}(t)+\kappa\dot{q}_{0}(t)+\omega_{0}^{2}q_{0}(t)=\underset{\mathrm{initial\,velocity\,change}}{\underbrace{-\kappa q_{0}(0)\delta(t)}}+\underset{\mathrm{stochastic\,force}}{\underbrace{\sum_{n}\frac{C_{n}}{m_{0}}\left[\dot{q}_{n}(0)\frac{\sin(\omega_{n}t)}{\omega_{n}}+q_{n}(0)\cos(\omega_{n}t)\right]}}.\label{eq:71}
\end{equation}
On the right-hand side, the second term presents the stochastic force
induced by the heat bath; the first term indicates an abrupt velocity
change $-\kappa q_{0}(0)$ of the system particle at the initial time
$t=0$ \citep{Bez1980,Canizares1994,Ju2017}. The sudden change of
velocity occurs for the system harmonic oscillator when the coupling
between the system and the heat bath is switched on. Such an initial
slippage is caused by the assumption of the initial product state.
To avoid such an initial discontinuous problem, we drop the first
term by considering the particle motion as starting at $t=0+$ \citep{Yu1994}.
Under such a modification, the Caldeira-Leggett model can reproduce
the complete Langevin equation with an arbitrary friction coefficient
$\kappa$ for both the underdamped and the overdamped regimes, and
the heat distribution of the Caldeira-Leggett model approaches that
of the classical Brownian motion described by the Kramers equation
\citep{Kramers1940}. In Appendix \ref{sec:Classical-Caldeira-Leggett-model},
for the classical counterpart of the Caldeira-Leggett model, we show
the initial slippage can be naturally eliminated by choosing another
initial state.

After dropping the first term, Eq. (\ref{eq:s^2+=00005Cetas}) becomes

\begin{equation}
(s^{2}+\kappa s+\omega_{0}^{2})\tilde{q}_{0}(s)=\dot{q}_{0}(0)+(\kappa+s)q_{0}(0)+\sum_{n}\frac{C_{n}}{m_{0}}\left[\frac{\dot{q}_{n}(0)+sq_{n}(0)}{s^{2}+\omega_{n}^{2}}\right].\label{eq:71-1}
\end{equation}
The solutions to $\tilde{q}_{0}(s)$ and $\tilde{q}_{n}(s)$ follow
immediately as

\begin{align}
\tilde{q}_{0}(s) & =\frac{\dot{q}_{0}(0)+(\kappa+s)q_{0}(0)+\sum_{n}\frac{C_{n}}{m_{0}}\frac{\dot{q}_{n}(0)+sq_{n}(0)}{s^{2}+\omega_{n}^{2}}}{s^{2}+\kappa s+\omega_{0}^{2}},\label{eq:qotildes}\\
\tilde{q}_{n}(s) & =\frac{\dot{q}_{n}(0)+sq_{n}(0)}{s^{2}+\omega_{n}^{2}}+\frac{C_{n}}{m_{n}}\cdot\frac{\dot{q}_{0}(0)+(\kappa+s)q_{0}(0)+\sum_{l}\frac{C_{l}}{m_{0}}\frac{\dot{q}_{l}(0)+sq_{l}(0)}{s^{2}+\omega_{l}^{2}}}{\left(s^{2}+\omega_{n}^{2}\right)\left(s^{2}+\kappa s+\omega_{0}^{2}\right)}.
\end{align}
With the inverse Laplace transform, the elements in the matrix $\exp\left(\mathbf{L}t\right)$
{[}Eq. (\ref{eq:expLt}){]} are determined by

\begin{equation}
\left(\begin{array}{c}
q_{0}(t)\\
q_{1}(t)\\
...\\
q_{N}(t)
\end{array}\right)=\left(\begin{array}{cccccccccc}
\alpha_{0} & \alpha_{1} & \alpha_{2} & ... & \alpha_{N} & \beta_{0} & \beta_{1} & \beta_{2} & ... & \beta_{N}\\
\gamma_{1} & \Lambda_{11} & \Lambda_{12} & ... & \Lambda_{1N} & \xi_{1} & \Delta_{11} & \Delta_{12} & ... & \Delta_{1N}\\
... & ... &  & ... &  & ... & ... &  & ...\\
\gamma_{N} & \Lambda_{N1} & \Lambda_{N2} & ... & \Lambda_{NN} & \xi_{N} & \Delta_{N1} & \Delta_{N2} & ... & \Delta_{NN}
\end{array}\right)\left(\begin{array}{c}
q_{0}(0)\\
q_{1}(0)\\
...\\
q_{N}(0)\\
\dot{q}_{0}(0)\\
\dot{q}_{1}(0)\\
...\\
\dot{q}_{N}(0)
\end{array}\right),
\end{equation}
where the elements in the matrix of the right-hand side are explicitly
solved as \citep{Yu1994} 
\begin{align}
\alpha_{0} & =e^{-\frac{\kappa t}{2}}\left[\cos(\hat{\omega}_{0}t)+\frac{\kappa}{2\hat{\omega}_{0}}\sin(\hat{\omega}_{0}t)\right],\\
\beta_{0} & =\frac{e^{-\frac{\kappa t}{2}}}{\hat{\omega}_{0}}\sin(\hat{\omega}_{0}t),\\
\alpha_{n} & =\frac{C_{n}}{m_{0}}f_{n}(t),\\
\beta_{n} & =\frac{C_{n}}{m_{0}}g_{n}(t),\\
\gamma_{n} & =\frac{C_{n}}{m_{n}}\left[f_{n}(t)+\kappa g_{n}(t)\right],\\
\xi_{n} & =\frac{C_{n}}{m_{n}}g_{n}(t),\\
\Lambda_{nl} & =\delta_{nl}\cos(\omega_{n}t)+\frac{C_{n}C_{l}}{m_{n}m_{0}}F_{nl}(t),\\
\Delta_{nl} & =\frac{\delta_{nl}}{\omega_{n}}\sin(\omega_{n}t)+\frac{C_{n}C_{l}}{m_{n}m_{0}}G_{nl}(t).
\end{align}
The functions $f_{n}(t),\,g_{n}(t),\,F_{nl}(t)$ and $G_{nl}(t)$
are explicitly

\begin{align}
f_{n}(t) & =\mathscr{L}^{-1}\left[\frac{s}{(s^{2}+\kappa s+\omega_{0}^{2})(s^{2}+\omega_{n}^{2})}\right],\\
g_{n}(t) & =\mathscr{L}^{-1}\left[\frac{1}{(s^{2}+\kappa s+\omega_{0}^{2})(s^{2}+\omega_{n}^{2})}\right],\\
F_{nl}(t) & =\mathscr{L}^{-1}\left[\frac{s}{(s^{2}+\kappa s+\omega_{0}^{2})(s^{2}+\omega_{n}^{2})(s^{2}+\omega_{l}^{2})}\right],\\
G_{nl}(t) & =\mathscr{L}^{-1}\left[\frac{1}{(s^{2}+\kappa s+\omega_{0}^{2})(s^{2}+\omega_{n}^{2})(s^{2}+\omega_{l}^{2})}\right],
\end{align}
where $\mathscr{L}^{-1}(\cdot)$ denotes the inverse Laplace transform
with $\mathscr{L}(\cdot)=\int_{0}^{\infty}(\cdot)e^{-st}dt$.

\subsection{$P(\mathbf{z})$}

$P(\mathbf{z})$ is the Wigner function of the state $\eta(0)$ for
the composite system \citep{Ford2001,Qiu2021}

\begin{equation}
P(\mathbf{z})=\frac{2\sinh\left(\frac{\beta^{\prime}\hbar\omega_{0}}{2}\right)}{\cosh\left[\frac{(\beta^{\prime}+i\nu)\hbar\omega_{0}}{2}\right]}\cdot\left[\prod_{n=1}^{N}2\tanh\left(\frac{\beta\hbar\omega_{n}}{2}\right)\right]\cdot\exp\left[-\frac{1}{2\hbar}\mathbf{z}^{\mathrm{T}}\boldsymbol{\Lambda}_{\beta z}\mathbf{z}\right],
\end{equation}
where $\boldsymbol{\Lambda}_{\beta z}$ is a $(2N+2)\times(2N+2)$
diagonal matrix

\begin{equation}
\boldsymbol{\Lambda}_{\beta z}=\mathrm{diag}(\lambda_{\beta^{\prime}q_{0}},\theta_{1},...,\theta_{N},\lambda_{\beta^{\prime}p_{0}},\mu_{1},...,\mu_{N}),\label{eq:Lambda_=00005Cbetaz}
\end{equation}
with the elements

\begin{align}
\theta_{n} & =2m_{n}\omega_{n}\tanh\left(\frac{\beta\hbar\omega_{n}}{2}\right),\\
\mu_{n} & =\frac{2}{m_{n}\omega_{n}}\tanh\left(\frac{\beta\hbar\omega_{n}}{2}\right),\\
\lambda_{\beta^{\prime}q_{0}} & =2m_{0}\omega_{0}\tanh\left[\frac{(\beta^{\prime}+i\nu)\hbar\omega_{0}}{2}\right],\\
\lambda_{\beta^{\prime}p_{0}} & =\frac{2}{m_{0}\omega_{0}}\tanh\left[\frac{(\beta^{\prime}+i\nu)\hbar\omega_{0}}{2}\right].
\end{align}

\subsection{Calculation of the integral\label{subsec:Calculation-of-the}}

With the explicit expressions of $\left[e^{i\nu H_{S}^{\mathrm{H}}(\tau)}\right]_{w}(\mathbf{z})$
and $P(\mathbf{z})$, we perform the integral in Eq. (\ref{eq:chiQ}),
and obtain the result of the characteristic function of heat 
\begin{align}
\chi_{\tau}(\nu) & =\sqrt{\frac{\det\left(\boldsymbol{\Lambda}_{\beta z}-i\boldsymbol{\Lambda}_{\nu z}\right)}{\det\left[\boldsymbol{\Lambda}_{\beta z}-i\boldsymbol{\tilde{\Lambda}}_{\nu z}(\tau)\right]}}.\label{eq:16-1}
\end{align}
We have used the following integral formula

\begin{equation}
\int d\mathbf{x}e^{-\frac{1}{2}\mathbf{x}^{\mathrm{T}}\mathbf{T}\mathbf{x}}=\sqrt{\frac{\left(2\pi\right)^{\dim\left(\mathbf{T}\right)}}{\det\left(\mathbf{T}\right)}},\label{eq:integral_formula}
\end{equation}
where all the eigenvalues of $\mathbf{T}$ have positive real parts.

By introducing a diagonal matrix $\mathbf{A}=\boldsymbol{\Lambda}_{\beta z}-i\boldsymbol{\Lambda}_{\nu z}$,
we rewrite Eq. (\ref{eq:16-1}) as

\begin{align}
\chi_{\tau}(\nu) & =\sqrt{\frac{1}{\det\left(\mathbf{I}+i\sqrt{\mathbf{A}^{-1}}\left[\boldsymbol{\Lambda}_{\nu z}-\boldsymbol{\tilde{\Lambda}}_{\nu z}(\tau)\right]\sqrt{\mathbf{A}^{-1}}\right)}}.\label{eq:67}
\end{align}
Since $\boldsymbol{\tilde{\Lambda}}_{\nu z}(\tau)$ is a rank-2 matrix,
we rewrite it as

\begin{equation}
\boldsymbol{\tilde{\Lambda}}_{\nu z}(\tau)=2m_{0}\omega_{0}\tan\left(\frac{\omega_{0}\hbar\nu}{2}\right)\left[\mathbf{v}_{q_{0}}(\tau)\mathbf{v}_{q_{0}}^{\mathrm{T}}(\tau)+\frac{1}{m_{0}^{2}\omega_{0}^{2}}\mathbf{v}_{p_{0}}(\tau)\mathbf{v}_{p_{0}}^{\mathrm{T}}(\tau)\right],\label{eq:Lambda_tilde_=00005Cnu_z}
\end{equation}
with the vectors

\begin{align}
\mathbf{v}_{q_{0}}(\tau) & =\left(\alpha_{0},\alpha_{1},...,\alpha_{N},\frac{\beta_{0}}{m_{0}},\frac{\beta_{1}}{m_{1}},...\frac{\beta_{N}}{m_{N}}\right)^{\mathrm{T}},\\
\mathbf{v}_{p_{0}}(\tau) & =\left(m_{0}\dot{\alpha}_{0},m_{0}\dot{\alpha}_{1},...,m_{0}\dot{\alpha}_{N},\dot{\beta}_{0},\frac{m_{0}}{m_{1}}\dot{\beta}_{1},...,\frac{m_{0}}{m_{N}}\dot{\beta}_{N}\right)^{\mathrm{T}}.
\end{align}
Here, the evolution time $t$ in the terms $\alpha_{n}$ and $\beta_{n}$
is set to $\tau$. We rewrite the matrix in the determinant {[}see
Eq. (\ref{eq:67}){]} as

\begin{equation}
\sqrt{\mathbf{A}^{-1}}\left(\boldsymbol{\Lambda}_{\nu z}-\boldsymbol{\tilde{\Lambda}}_{\nu z}(\tau)\right)\sqrt{\mathbf{A}^{-1}}=\mathbf{M}\mathbf{M}^{\mathrm{T}},
\end{equation}
with the matrix

\begin{equation}
\mathbf{M}^{\mathrm{T}}=\left(\begin{array}{c}
\sqrt{2m_{0}\omega_{0}\tan\frac{\omega_{0}\hbar\nu}{2}}\mathbf{v}_{q_{0}}^{\mathrm{T}}(0)\\
\sqrt{\frac{2}{m_{0}\omega_{0}}\tan\frac{\omega_{0}\hbar\nu}{2}}\mathbf{v}_{p_{0}}^{\mathrm{T}}(0)\\
i\sqrt{2m_{0}\omega_{0}\tan\frac{\omega_{0}\hbar\nu}{2}}\mathbf{v}_{q_{0}}^{\mathrm{T}}(\tau)\\
i\sqrt{\frac{2}{m_{0}\omega_{0}}\tan\frac{\omega_{0}\hbar\nu}{2}}\mathbf{v}_{p_{0}}^{\mathrm{T}}(\tau)
\end{array}\right)\sqrt{\mathbf{A}^{-1}}.
\end{equation}

The determinant in Eq. (\ref{eq:67}) can be simplified to

\begin{equation}
\det\left(\mathbf{I}+i\sqrt{\mathbf{A}^{-1}}\left[\boldsymbol{\Lambda}_{\nu z}-\boldsymbol{\tilde{\Lambda}}_{\nu z}(\tau)\right]\sqrt{\mathbf{A}^{-1}}\right)=\det\left(\mathbf{I}_{4}+i\mathbf{M}^{\mathrm{T}}\mathbf{M}\right),\label{eq:determinant_4times4}
\end{equation}
where the right-hand side is the determinant of a $4\times4$ matrix,
and $\mathbf{I}_{4}$ is the $4\times4$ identity matrix. Notice that
the initial values of the two vectors are

\begin{align}
\mathbf{v}_{q_{0}}(0) & =\left(1,0,...0,0,0,...,0\right)^{\mathrm{T}},\\
\mathbf{v}_{p_{0}}(0) & =\left(0,0,...0,1,0,...,0\right)^{\mathrm{T}}.
\end{align}
The explicit result of $\mathbf{M}^{\mathrm{T}}\mathbf{M}$ is obtained
as

\begin{equation}
\mathbf{M}^{\mathrm{T}}\mathbf{M}=\Xi\left(\begin{array}{cccc}
1 & 0 & i\alpha_{0} & i\dot{\alpha}_{0}/\omega_{0}\\
0 & 1 & i\omega_{0}\beta_{0} & i\dot{\beta}_{0}\\
i\alpha_{0} & i\omega_{0}\beta_{0} & -h_{11}(\tau) & -h_{12}(\tau)\\
i\dot{\alpha}_{0}/\omega_{0} & i\dot{\beta}_{0} & -h_{12}(\tau) & -h_{22}(\tau)
\end{array}\right),\label{eq:N_matrix}
\end{equation}
where the elements are functions of the final time $\tau$, and the
functions $h_{11}(\tau)$, $h_{12}(\tau)$ and $h_{22}(\tau)$ are

\begin{align}
h_{11}(\tau) & =\frac{2m_{0}\omega_{0}}{\Xi}\tan\left(\frac{\omega_{0}\hbar\nu}{2}\right)\mathbf{v}_{q_{0}}^{\mathrm{T}}(\tau)\mathbf{A}^{-1}\mathbf{v}_{q_{0}}(\tau),\\
h_{22}(\tau) & =\frac{2}{m_{0}\omega_{0}\Xi}\tan\left(\frac{\omega_{0}\hbar\nu}{2}\right)\mathbf{v}_{p_{0}}^{\mathrm{T}}(\tau)\mathbf{A}^{-1}\mathbf{v}_{p_{0}}(\tau),\\
h_{12}(\tau) & =\frac{2}{\Xi}\tan\left(\frac{\omega_{0}\hbar\nu}{2}\right)\mathbf{v}_{q_{0}}^{\mathrm{T}}(\tau)\mathbf{A}^{-1}\mathbf{v}_{p_{0}}(\tau),
\end{align}
with

\begin{align}
\mathbf{v}_{q_{0}}^{\mathrm{T}}(t)\mathbf{A}^{-1}\mathbf{v}_{q_{0}}(t) & =\frac{\alpha_{0}^{2}+\omega_{0}^{2}\beta_{0}^{2}}{2m_{0}\omega_{0}\left\{ \tanh\left[\frac{(\beta^{\prime}+i\nu)\hbar\omega_{0}}{2}\right]-i\tan\left(\frac{\omega_{0}\hbar\nu}{2}\right)\right\} }+\sum_{n=1}^{N}\left(\frac{\alpha_{n}^{2}}{\theta_{n}}+\frac{1}{\mu_{n}}\frac{\beta_{n}^{2}}{m_{n}^{2}}\right),\\
\mathbf{v}_{p_{0}}^{\mathrm{T}}(t)\mathbf{A}^{-1}\mathbf{v}_{p_{0}}(t) & =\frac{m_{0}\left(\dot{\alpha}_{0}^{2}+\omega_{0}^{2}\dot{\beta}_{0}^{2}\right)}{2\omega_{0}\left\{ \tanh\left[\frac{(\beta^{\prime}+i\nu)\hbar\omega_{0}}{2}\right]-i\tan\left(\frac{\omega_{0}\hbar\nu}{2}\right)\right\} }+m_{0}^{2}\sum_{n=1}^{N}\left(\frac{\dot{\alpha}_{n}^{2}}{\theta_{n}}+\frac{1}{\mu_{n}}\frac{\dot{\beta}_{n}^{2}}{m_{n}^{2}}\right),\\
\mathbf{v}_{q_{0}}^{\mathrm{T}}(t)\mathbf{A}^{-1}\mathbf{v}_{p_{0}}(t) & =\frac{\frac{d}{dt}\left(\alpha_{0}^{2}+\omega_{0}^{2}\beta_{0}^{2}\right)}{4\omega_{0}\left\{ \tanh\left[\frac{(\beta^{\prime}+i\nu)\hbar\omega_{0}}{2}\right]-i\tan\left(\frac{\omega_{0}\hbar\nu}{2}\right)\right\} }+\frac{m_{0}}{2}\sum_{n=1}^{N}\frac{d}{dt}\left(\frac{\alpha_{n}^{2}}{\theta_{n}}+\frac{1}{\mu_{n}}\frac{\beta_{n}^{2}}{m_{n}^{2}}\right).
\end{align}
The summations are replaced by the integral with the Ohmic spectral
density, and every element in Eq. (\ref{eq:N_matrix}) is carried
out as

\begin{align}
\alpha_{0}(\tau) & =e^{-\frac{\kappa\tau}{2}}\left[\cos\left(\hat{\omega}_{0}\tau\right)+\frac{\kappa\sin\left(\hat{\omega}_{0}\tau\right)}{2\hat{\omega}_{0}}\right],\\
\beta_{0}(\tau) & =\frac{e^{-\frac{\kappa\tau}{2}}\sin\left(\hat{\omega}_{0}\tau\right)}{\hat{\omega}_{0}},\\
h_{11}(\tau) & =\Theta+e^{-\kappa\tau}\left[\frac{\omega_{0}^{2}}{\hat{\omega}_{0}^{2}}+\frac{\kappa\sin\left(2\hat{\omega}_{0}\tau\right)}{2\hat{\omega}_{0}}-\frac{\kappa^{2}\cos\left(2\hat{\omega}_{0}\tau\right)}{4\hat{\omega}_{0}^{2}}\right]\left(1-\Theta\right),\\
h_{22}(\tau) & =\Theta+e^{-\kappa\tau}\left[\frac{\omega_{0}^{2}}{\hat{\omega}_{0}^{2}}-\frac{\kappa\sin\left(2\hat{\omega}_{0}\tau\right)}{2\hat{\omega}_{0}}-\frac{\kappa^{2}\cos\left(2\hat{\omega}_{0}\tau\right)}{4\hat{\omega}_{0}^{2}}\right]\left(1-\Theta\right),\\
h_{12}(\tau) & =\frac{\kappa\omega_{0}e^{-\kappa\tau}}{2\hat{\omega}_{0}^{2}}\left(\Theta-1\right)\left[1-\cos\left(2\hat{\omega}_{0}\tau\right)\right].\label{eq:elements_in_mat_N(tau)}
\end{align}
Then, Eq. (\ref{eq:characteristic_function_CL_model}) is obtained
by directly calculating the determinant of a $4\times4$ matrix in
Eq. (\ref{eq:determinant_4times4}).

\section{Classical Caldeira-Leggett model\label{sec:Classical-Caldeira-Leggett-model}}

We consider the classical Caldeira-Leggett model, where coordinates
and momenta commute with each other. To eliminate the initial slippage,
the initial state is amended as a coupled state
\begin{align}
\rho^{\mathrm{cl}}(\mathbf{z};0) & =\frac{e^{-\beta^{\prime}H_{S}(0)-\beta[H_{B}(0)+H_{SB}(0)]}}{Z^{\mathrm{cl}}(\beta^{\prime},\beta)},\label{eq:81}
\end{align}
which represents the probability density in the phase space of the
composite system. The classical partition function is obtained by
performing the integral in the phase space

\begin{align}
Z^{\mathrm{cl}}(\beta^{\prime},\beta) & =\iint e^{-\beta^{\prime}H_{S}(0)-\beta[H_{B}(0)+H_{SB}(0)]}dq_{0}dq_{1}...dq_{N}dp_{0}dp_{1}...dp_{N}\\
 & =\frac{2\pi}{\beta^{\prime}\omega_{0}}\prod_{n=1}^{N}\left(\frac{2\pi}{\beta\omega_{n}}\right),
\end{align}
which is independent of the interaction (notice that the partition
function of the quantum model relies on the interaction strength \citep{Grabert1984,Weiss2008}).

We also define the classical fluctuating heat as the energy difference
of the initial and the final system energy. For classical dynamics,
the initial and the final states are directly represented by the points
in the phase space, and the measurements over the system can be applied
without disturbing the composite system. Therefore, the characteristic
function of heat is 
\begin{equation}
\chi_{\tau}^{\mathrm{cl}}(\nu)=\frac{\iint e^{i\nu H_{S}(\tau)-(\beta^{\prime}+i\nu)H_{S}(0)-\beta[H_{B}(0)+H_{SB}(0)]}dq_{0}dq_{1}...dq_{N}dp_{0}dp_{1}...dp_{N}}{Z^{\mathrm{cl}}(\beta^{\prime},\beta)},\label{eq:84}
\end{equation}
where the energy of the system $H_{S}(t)=[p_{0}(t)]^{2}/(2m_{0})+m_{0}\omega_{0}^{2}[q_{0}(t)]^{2}/2$
is determined by $p_{0}(t)$ and $q_{0}(t)$ associated with the initial
point $\mathbf{z}$. We choose a new set of initial variables $q_{0}$,
$\mathfrak{q}_{n}\coloneqq q_{n}-C_{n}q_{0}/(m_{n}\omega_{n}^{2})$,
$p_{0}$ and $p_{n}$ in the following calculation.

We rewrite the evolution of the coordinate $q_{0}(t)$ of the system
{[}Eq. (\ref{eq:s^2+=00005Cetas}){]} as 
\begin{equation}
(s^{2}+\kappa s+\omega_{0}^{2})\tilde{q}_{0}(s)=\dot{q}_{0}(0)+\left(s+\sum_{n}\frac{C_{n}^{2}}{m_{0}m_{n}\omega_{n}^{2}}\frac{s}{s^{2}+\omega_{n}^{2}}\right)q_{0}(0)+\sum_{n}\frac{C_{n}}{m_{0}}\frac{\dot{q}_{n}(0)+s\mathfrak{q}_{n}}{s^{2}+\omega_{n}^{2}}.\label{eq:s^2+=00005Cetas-1}
\end{equation}
For the Ohmic spectral density, the summation in the second term is
\begin{equation}
\sum_{n}\frac{C_{n}^{2}}{m_{0}m_{n}\omega_{n}^{2}}\frac{s}{s^{2}+\omega_{n}^{2}}=\kappa.
\end{equation}
Thus, Eq. (\ref{eq:s^2+=00005Cetas-1}) naturally leads to Eq. (\ref{eq:71-1})
by substituting $q_{n}(0)$ into $\mathfrak{q}_{n}$. The initial
slippage is rationally eliminated by choosing a coupled initial state
{[}Eq. (\ref{eq:81}){]}. In reality, the interaction between the
system and the heat bath always exist, and one cannot prepare the
initial state of the composite system without the influence of the
interaction. The initial state of the composite system is more likely
in the coupled form {[}Eq. (\ref{eq:81}){]}. The heat bath encodes
partial information of the system due to the interaction.

Similar to Eq. (\ref{eq:33}), the system energy at time $t$ can
be represented by the dynamical map as

\begin{align}
H_{S}(t) & =\frac{1}{2}\mathbf{z}^{\mathrm{T}}(t)\mathbf{\Lambda}_{H_{S}}\mathbf{z}(t)\\
 & =\frac{1}{2}\mathbf{z}^{\mathrm{T}}(0)\tilde{\mathbf{\Lambda}}_{H_{S}}(t)\mathbf{z}(0),
\end{align}
with the matrices \footnote{Strictly, the substitution requires to amend $\gamma_{n},\Lambda_{nm},\xi_{n},\Delta_{nm}$
accordingly, but we only require $q_{0}(t)$ and $p_{0}(t)$ to calculate
the characteristic function of heat, so we skip the further amendment.}

\begin{align}
\mathbf{\Lambda}_{H_{S}} & =\left(\begin{array}{cccc}
m_{0}\omega_{0}^{2}\\
 & \mathbf{O}\\
 &  & \frac{1}{m_{0}}\\
 &  &  & \mathbf{O}
\end{array}\right),
\end{align}
and 
\begin{equation}
\tilde{\mathbf{\Lambda}}_{H_{S}}(t)=\exp(\mathbf{L}^{\mathrm{T}}t)\mathbf{\Lambda}_{H_{S}}\exp(\mathbf{L}t).
\end{equation}
The initial vector is now amended to

\begin{equation}
\mathbf{z}(0)=\left(q_{0}(0),\mathfrak{q}_{1},...,\mathfrak{q}_{n},p_{0}(0),...,p_{N}(0)\right)^{\mathrm{T}}.
\end{equation}
The initial Hamiltonians $H_{S}(0)$ and $H_{B}(0)+H_{SB}(0)$ are
\begin{align}
H_{S}(0) & =\frac{1}{2}\mathbf{z}^{\mathrm{T}}(0)\mathbf{\Lambda}_{H_{S}}\mathbf{z}(0),\\
H_{B}(0)+H_{SB}(0) & =\sum_{n=1}^{N}\left(\frac{1}{2}\frac{p_{n}^{2}}{m_{n}}+\frac{1}{2}m_{n}\omega_{n}^{2}\mathfrak{q}_{n}^{2}\right)\nonumber \\
 & =\frac{1}{2}\mathbf{z}^{\mathrm{T}}(0)\mathbf{\Lambda}_{H_{B}}\mathbf{z}(0),
\end{align}
with the matrix
\begin{equation}
\mathbf{\Lambda}_{H_{B}}=\mathrm{diag}(0,m_{1}\omega_{1}^{2},...,m_{N}\omega_{N}^{2},0,\frac{1}{m_{1}},...,\frac{1}{m_{N}}).
\end{equation}

According to the integral formula (\ref{eq:integral_formula}), we
carry out the characteristic function Eq. (\ref{eq:84}) into

\begin{equation}
\chi_{\tau}^{\mathrm{cl}}(\nu)=\sqrt{\frac{\det\left[\beta^{\prime}\mathbf{\Lambda}_{H_{S}}+\beta\mathbf{\Lambda}_{H_{B}}\right]}{\det\left[\beta^{\prime}\mathbf{\Lambda}_{H_{S}}+\beta\mathbf{\Lambda}_{H_{B}}-i\nu\left(\tilde{\mathbf{\Lambda}}_{H_{S}}(\tau)-\mathbf{\Lambda}_{H_{S}}\right)\right]}}.\label{eq:chitauclassical}
\end{equation}
For the classical limit $\hbar\rightarrow0$, we can verify

\begin{align}
\lim_{\hbar\rightarrow0}\frac{\mathbf{\Lambda}_{\beta z}}{\hbar} & =(\beta^{\prime}+i\nu)\mathbf{\Lambda}_{H_{S}}+\beta\mathbf{\Lambda}_{H_{B}},\\
\lim_{\hbar\rightarrow0}\frac{\mathbf{\Lambda}_{\nu z}}{\hbar} & =\nu\mathbf{\Lambda}_{H_{S}},\\
\lim_{\hbar\rightarrow0}\frac{\tilde{\mathbf{\Lambda}}_{\nu z}(t)}{\hbar} & =\nu\tilde{\mathbf{\Lambda}}_{H_{S}}(t),
\end{align}
and obtain
\begin{equation}
\lim_{\hbar\rightarrow0}\chi_{\tau}(\nu)=\chi_{\tau}^{\mathrm{cl}}(\nu),
\end{equation}
with $\chi_{\tau}(\nu)$ given in Eq. (\ref{eq:16-1}). The final
result {[}Eq. (\ref{eq:chitauclassical}){]} is the same as Eq. (\ref{eq:characteristic_function_classical_result}).
Hence, we use the same notation.

\section{The characteristic function of heat for the classical Brownian motion\label{sec:characteristic-function-of}}

We derive the characteristic function of heat for the classical Brownian
motion. For an underdamped Brownian particle moving in a potential
$V(x)$, the stochastic dynamics is described by the complete Langevin
equation

\begin{equation}
\ddot{x}+\kappa\dot{x}+\frac{1}{m}\frac{\partial V}{\partial x}=\frac{1}{m}F_{\mathrm{fluc}}(t).
\end{equation}
The fluctuating force is a Gaussian white noise satisfying the fluctuation-dissipation
relation

\begin{equation}
\left\langle F_{\mathrm{fluc}}(t)F_{\mathrm{fluc}}(t^{\prime})\right\rangle =2m\kappa k_{B}T\delta(t-t^{\prime}).
\end{equation}
The evolution of the system state is characterized by the probability
density function $\rho(x,p;t)$ in the phase space. The stochastic
dynamics is then described by the Kramers equation \citep{Kramers1940}
\begin{equation}
\frac{\partial\rho}{\partial t}=\mathscr{L}[\rho],\label{eq:differential_equation_Kramersss}
\end{equation}
with the Liouville operator

\begin{equation}
\mathscr{L}[\rho]=-\frac{\partial}{\partial x}(\frac{p}{m}\rho)+\frac{\partial}{\partial p}\left[\kappa p\rho+\frac{\partial V(x)}{\partial x}\rho+\frac{\kappa m}{\beta}\frac{\partial\rho}{\partial p}\right].\label{eq:Lrho}
\end{equation}

Similarly in the phase space, we calculate the characteristic function
of heat for the classical Brownian motion

\begin{align}
\chi_{\tau}^{\mathrm{cl}}(\nu) & =\iint dxdpe^{i\nu\left[\frac{p^{2}}{2m}+V(x)\right]}\eta(x,p;\tau),\label{eq:chitauclassical_quadraticform}
\end{align}
where a probability-density-like function $\eta(x,p;t)$ also satisfies
the dynamic equation (\ref{eq:differential_equation_Kramersss}) with
the initial condition

\begin{equation}
\eta(x,p;0)=e^{-i\nu\left[\frac{p^{2}}{2m}+V(x)\right]}\rho(x,p;0).
\end{equation}

We consider the system potential as a harmonic potential $V(x)=m\omega_{0}^{2}x^{2}/2$
and the initial system state as an equilibrium state

\begin{equation}
\rho(x,p;0)=\frac{1}{Z_{S}^{\mathrm{cl}}(\beta^{\prime})}e^{-\beta^{\prime}\left(\frac{p^{2}}{2m}+\frac{1}{2}m\omega_{0}^{2}x^{2}\right)},
\end{equation}
with the inverse temperature $\beta^{\prime}$ and the classical partition
function $Z_{S}^{\mathrm{cl}}(\beta^{\prime})=2\pi/(\beta^{\prime}\omega_{0})$.
Under such conditions, the probability-density-like function $\eta(x,p;t)$
is always in a quadratic form, assumed as

\begin{equation}
\eta(x,p;t)=\frac{1}{Z_{S}^{\mathrm{cl}}(\beta^{\prime})}e^{-\left[a(t)\frac{p^{2}}{2m}+b(t)\frac{1}{2}m\omega_{0}^{2}x^{2}+c(t)\omega_{0}xp+\Lambda(t)\right]}.\label{eq:quadraticform}
\end{equation}
The Kramers equation (\ref{eq:differential_equation_Kramersss}) for
$\eta(x,p;t)$ leads to the following ordinary differential equations

\begin{align}
\dot{\Lambda} & =-\kappa\left(1-\frac{a}{\beta}\right),\\
\dot{a} & =2\kappa a\left(1-\frac{a}{\beta}\right)-2\omega_{0}c,\label{eq:93}\\
\dot{b} & =2c\left(\omega_{0}-\frac{\kappa}{\beta}c\right),\label{eq:94}\\
\dot{c} & =\omega_{0}\left(a-b\right)+\kappa c-2\frac{\kappa}{\beta}ac,\label{eq:95}
\end{align}
with the initial conditions $a(0)=b(0)=\beta^{\prime}+i\nu$, $c(0)=0$
and $\Lambda(0)=0$. According to the conservation of the probability
$\iint\eta(x,p;t)dxdp=\mathrm{const}$, the coefficient $\Lambda(t)$
is obtained as

\begin{align}
e^{-\Lambda(t)} & =\frac{\sqrt{a(t)b(t)-c(t)^{2}}}{\beta^{\prime}+i\nu}.
\end{align}
Substituting Eq. (\ref{eq:quadraticform}) into Eq. (\ref{eq:chitauclassical_quadraticform}),
we obtain the characteristic function for the classical Brownian motion
as

\begin{equation}
\chi_{\tau}^{\mathrm{cl}}(\nu)=\frac{\beta^{\prime}}{\beta^{\prime}+i\nu}\sqrt{\frac{a(\tau)b(\tau)-c(\tau)^{2}}{[a(\tau)-i\nu][b(\tau)-i\nu]-c(\tau)^{2}}}.\label{eq:chiQbetter_result}
\end{equation}

To solve the nonlinear differential equations (\ref{eq:93})-(\ref{eq:95}),
we introduce a new set of variables

\begin{align}
A & =\frac{a}{ab-c^{2}},\\
B & =\frac{b}{ab-c^{2}},\\
C & =\frac{c}{ab-c^{2}},
\end{align}
and obtain the linear differential equations

\begin{align}
\frac{dA}{dt} & =-2\omega_{0}C,\label{eq:dA/dt}\\
\frac{dB}{dt} & =2\omega_{0}C-2\kappa B+2\frac{\kappa}{\beta},\\
\frac{dC}{dt} & =\omega_{0}(A-B)-\kappa C,\label{eq:dC/dt}
\end{align}
with the initial conditions $A(0)=B(0)=1/(\beta^{\prime}+i\nu)$ and
$C(0)=0$. The characteristic function Eq. (\ref{eq:chiQbetter_result})
becomes

\begin{equation}
\chi_{\tau}^{\mathrm{cl}}(\nu)=\frac{\beta^{\prime}}{\beta^{\prime}+i\nu}\sqrt{\frac{1}{1-i\nu(A+B)-\nu^{2}(AB-C^{2})}}.\label{eq:121}
\end{equation}
The solutions to Eqs. (\ref{eq:dA/dt})-(\ref{eq:dC/dt}) are

\begin{align}
A(t) & =-\frac{e^{-\kappa t}\left[\kappa^{2}\cos\left(2\hat{\omega}_{0}t\right)-2\hat{\omega}_{0}\kappa\sin\left(2\hat{\omega}_{0}t\right)-4\omega_{0}^{2}\right]}{4\beta\hat{\omega}_{0}^{2}}\frac{\beta-\beta^{\prime}-i\nu}{\beta^{\prime}+i\nu}+\frac{1}{\beta},\\
B(t) & =-\frac{e^{-\kappa t}\left[\kappa^{2}\cos\left(2\hat{\omega}_{0}t\right)+2\hat{\omega}_{0}\kappa\sin\left(2\hat{\omega}_{0}t\right)-4\omega_{0}^{2}\right]}{4\beta\hat{\omega}_{0}^{2}}\frac{\beta-\beta^{\prime}-i\nu}{\beta^{\prime}+i\nu}+\frac{1}{\beta},\\
C(t) & =-\frac{2e^{-\kappa t}\kappa\omega_{0}\left[\cos\left(2\hat{\omega}_{0}t\right)-1\right]}{4\beta\hat{\omega}_{0}^{2}}\frac{\beta-\beta^{\prime}-i\nu}{\beta^{\prime}+i\nu},
\end{align}
with $\hat{\omega}_{0}=\sqrt{\omega_{0}^{2}-\kappa^{2}/4}$. Plugging
the solutions into Eq. (\ref{eq:121}), we immediately obtain Eq.
(\ref{eq:characteristic_function_classical_result}). We remark that
the heat distribution of the classical Brownian motion has been obtained
by the path-integral method in Ref. \citep{Paraguassu2021a}, but
they only consider the initial temperature of the system to be the
same as that of the bath.

\subsection{The long-time limit}

After sufficiently long relaxation time, the solutions $a(t)$, $b(t)$
and $c(t)$ to Eqs. (\ref{eq:93})-(\ref{eq:95}) eventually approach
$a(\infty)=b(\infty)=\beta$ and $c(\infty)=0$. The long-time limit
of Eq. (\ref{eq:chiQbetter_result}) reproduces Eq. (\ref{eq:chiinfclassical}).

\subsection{The underdamped limit}

In the underdamped limit $\kappa/\omega_{0}\rightarrow0$, the differential
equations (\ref{eq:93})-(\ref{eq:95}) are reduced to

\begin{equation}
\dot{a}=\kappa a\left(1-\frac{a}{\beta}\right),
\end{equation}
with $b=a$ and $c=0$. The solution is

\begin{equation}
a(t)=\frac{\beta(\beta^{\prime}+i\nu)}{\beta^{\prime}+i\nu+(\beta-\beta^{\prime}-i\nu)e^{-\kappa t}},
\end{equation}
and Eq. (\ref{eq:chiQbetter_result}) becomes

\begin{equation}
\chi_{\tau}^{\mathrm{w},\mathrm{cl}}(\nu)=\frac{\beta\beta^{\prime}}{(\beta-i\nu)(\beta^{\prime}+i\nu)(1-e^{-\kappa\tau})+\beta\beta^{\prime}e^{-\kappa\tau}}.\label{eq:classical_chi_Q}
\end{equation}
It can be checked that Eq. (\ref{eq:weak_characteristic_function})
reproduces Eq. (\ref{eq:classical_chi_Q}) in the classical limit
$\hbar\rightarrow0$.

\subsection{The overdamped limit}

In the overdamped limit $\kappa/\omega_{0}\rightarrow\infty$, the
differential equations (\ref{eq:93})-(\ref{eq:95}) are reduced to
\begin{align}
\dot{a} & =2\kappa a\left(1-\frac{a}{\beta}\right),\\
\dot{b} & =2\omega_{0}c\left(1-\frac{\kappa}{\beta\omega_{0}}c\right),\label{eq:80}\\
0 & =\left(\kappa c+\omega_{0}a\right)-2\frac{\kappa}{\beta}ac-b\omega_{0}.\label{eq:57-1}
\end{align}
Eliminating $c$ in Eq. (\ref{eq:57-1}), Eq. (\ref{eq:80}) becomes

\begin{equation}
\dot{b}=\frac{2\omega_{0}^{2}}{\kappa}\frac{\left(\beta-a-b\right)\left(b-a\right)}{\left(\beta-2a\right)\left(1-2\frac{a}{\beta}\right)}.\label{eq:103}
\end{equation}
Notice that in the overdamped limit, the relaxation timescales of
the momentum and the coordinate are separated. We can substitute $a=\beta$
in Eq. (\ref{eq:103}) and obtain

\begin{equation}
\dot{b}=\frac{2\omega_{0}^{2}}{\kappa}b\left(1-\frac{b}{\beta}\right).
\end{equation}
With the initial condition $a(0)=\beta^{\prime}+i\nu$ and $b(0)=\beta^{\prime}+i\nu$,
the solutions are

\begin{align}
a(t) & =\frac{\left(\beta^{\prime}+i\nu\right)\beta}{\beta^{\prime}+i\nu+\left(\beta-\beta^{\prime}-i\nu\right)e^{-2\kappa t}},\label{eq:104}\\
b(t) & =\frac{\left(\beta^{\prime}+i\nu\right)\beta}{\beta^{\prime}+i\nu+\left(\beta-\beta^{\prime}-i\nu\right)e^{-\frac{2\omega_{0}^{2}}{\kappa}t}}.\label{eq:105}
\end{align}
We substitute Eqs. (\ref{eq:104}), (\ref{eq:105}) and $c(t)\approx0$
into Eq. (\ref{eq:chiQbetter_result}) and obtain

\begin{align}
\chi_{\tau}^{\mathrm{s},\mathrm{cl}}(\nu)= & \frac{\beta\beta^{\prime}}{\sqrt{(\beta-i\nu)(\beta^{\prime}+i\nu)\left(1-e^{-2\kappa\tau}\right)+\beta\beta^{\prime}e^{-2\kappa\tau}}}\nonumber \\
 & \times\frac{1}{\sqrt{(\beta-i\nu)(\beta^{\prime}+i\nu)\left(1-e^{-\frac{2\omega_{0}^{2}}{\kappa}\tau}\right)+\beta\beta^{\prime}e^{-\frac{2\omega_{0}^{2}}{\kappa}\tau}}}.\label{eq:strong_charateristic_function_classical}
\end{align}
It can be checked that Eq. (\ref{eq:strong_charateristic_function})
reproduces Eq. (\ref{eq:strong_charateristic_function_classical})
in the classical limit $\hbar\rightarrow0$.

\section{The characteristic function of heat for complete thermalization\label{sec:heat_characteristic_function_complete_thermalization}}

We derive the characteristic function of heat for a complete thermalization
process {[}Eq. (\ref{eq:chiQ_thermalization}){]} in the main content.
For a complete thermalization process (typically with infinite relaxation
time), the information of the initial state is completely forgotten,
and the final state is always an equilibrium state at the inverse
temperature $\beta$ of the heat bath

\begin{equation}
\gamma_{\mathrm{th},l^{\prime}l}=p_{l^{\prime}}^{\mathrm{eq}},
\end{equation}
regardless of the initial state $l$. Therefore, the characteristic
function of heat for complete thermalization is $\chi_{\mathrm{th}}(\nu)=\sum_{l^{\prime},l}\exp[i\nu(E_{l^{\prime}}^{S}-E_{l}^{S})]p_{l}p_{l^{\prime}}^{\mathrm{eq}}$.
We immediately obtain Eq. (\ref{eq:chiQ_thermalization}) by plugging
into the initial distribution $p_{l}=\exp(-\beta^{\prime}E_{l}^{S})/Z_{S}(\beta^{\prime})$
and the final distribution $p_{l^{\prime}}^{\mathrm{eq}}=\exp(-\beta E_{l^{\prime}}^{S})/Z_{S}(\beta)$,
where $Z_{S}(\beta^{\prime})=\sum_{l}\exp(-\beta^{\prime}E_{l}^{S})$
is the partition function of the system at the inverse temperature
$\beta^{\prime}$.

\bibliographystyle{apsrev4-1}
\bibliography{quantum_heat_ref}

\begin{thebibliography}{88}%
\makeatletter
\providecommand \@ifxundefined [1]{%
 \@ifx{#1\undefined}
}%
\providecommand \@ifnum [1]{%
 \ifnum #1\expandafter \@firstoftwo
 \else \expandafter \@secondoftwo
 \fi
}%
\providecommand \@ifx [1]{%
 \ifx #1\expandafter \@firstoftwo
 \else \expandafter \@secondoftwo
 \fi
}%
\providecommand \natexlab [1]{#1}%
\providecommand \enquote  [1]{``#1''}%
\providecommand \bibnamefont  [1]{#1}%
\providecommand \bibfnamefont [1]{#1}%
\providecommand \citenamefont [1]{#1}%
\providecommand \href@noop [0]{\@secondoftwo}%
\providecommand \href [0]{\begingroup \@sanitize@url \@href}%
\providecommand \@href[1]{\@@startlink{#1}\@@href}%
\providecommand \@@href[1]{\endgroup#1\@@endlink}%
\providecommand \@sanitize@url [0]{\catcode `\\12\catcode `\$12\catcode
  `\&12\catcode `\#12\catcode `\^12\catcode `\_12\catcode `\%12\relax}%
\providecommand \@@startlink[1]{}%
\providecommand \@@endlink[0]{}%
\providecommand \url  [0]{\begingroup\@sanitize@url \@url }%
\providecommand \@url [1]{\endgroup\@href {#1}{\urlprefix }}%
\providecommand \urlprefix  [0]{URL }%
\providecommand \Eprint [0]{\href }%
\providecommand \doibase [0]{http://dx.doi.org/}%
\providecommand \selectlanguage [0]{\@gobble}%
\providecommand \bibinfo  [0]{\@secondoftwo}%
\providecommand \bibfield  [0]{\@secondoftwo}%
\providecommand \translation [1]{[#1]}%
\providecommand \BibitemOpen [0]{}%
\providecommand \bibitemStop [0]{}%
\providecommand \bibitemNoStop [0]{.\EOS\space}%
\providecommand \EOS [0]{\spacefactor3000\relax}%
\providecommand \BibitemShut  [1]{\csname bibitem#1\endcsname}%
\let\auto@bib@innerbib\@empty
\bibitem [{\citenamefont {Gallavotti}\ and\ \citenamefont
  {Cohen}(1995)}]{Gallavotti1995a}%
  \BibitemOpen
  \bibfield  {author} {\bibinfo {author} {\bibfnamefont {G.}~\bibnamefont
  {Gallavotti}}\ and\ \bibinfo {author} {\bibfnamefont {E.~G.~D.}\ \bibnamefont
  {Cohen}},\ }\href {\doibase 10.1103/physrevlett.74.2694} {\bibfield
  {journal} {\bibinfo  {journal} {Phys. Rev. Lett.}\ }\textbf {\bibinfo
  {volume} {74}},\ \bibinfo {pages} {2694} (\bibinfo {year}
  {1995})}\BibitemShut {NoStop}%
\bibitem [{\citenamefont {Jarzynski}(1997)}]{Jarzynski1997}%
  \BibitemOpen
  \bibfield  {author} {\bibinfo {author} {\bibfnamefont {C.}~\bibnamefont
  {Jarzynski}},\ }\href {\doibase 10.1103/physrevlett.78.2690} {\bibfield
  {journal} {\bibinfo  {journal} {Phys. Rev. Lett.}\ }\textbf {\bibinfo
  {volume} {78}},\ \bibinfo {pages} {2690} (\bibinfo {year}
  {1997})}\BibitemShut {NoStop}%
\bibitem [{\citenamefont {Crooks}(1999)}]{Crooks1999}%
  \BibitemOpen
  \bibfield  {author} {\bibinfo {author} {\bibfnamefont {G.~E.}\ \bibnamefont
  {Crooks}},\ }\href {\doibase 10.1103/physreve.60.2721} {\bibfield  {journal}
  {\bibinfo  {journal} {Phys. Rev. E}\ }\textbf {\bibinfo {volume} {60}},\
  \bibinfo {pages} {2721} (\bibinfo {year} {1999})}\BibitemShut {NoStop}%
\bibitem [{\citenamefont {Jarzynski}\ and\ \citenamefont
  {W{\'{o}}jcik}(2004)}]{Jarzynski2004}%
  \BibitemOpen
  \bibfield  {author} {\bibinfo {author} {\bibfnamefont {C.}~\bibnamefont
  {Jarzynski}}\ and\ \bibinfo {author} {\bibfnamefont {D.~K.}\ \bibnamefont
  {W{\'{o}}jcik}},\ }\href {\doibase 10.1103/physrevlett.92.230602} {\bibfield
  {journal} {\bibinfo  {journal} {Phys. Rev. Lett.}\ }\textbf {\bibinfo
  {volume} {92}},\ \bibinfo {pages} {230602} (\bibinfo {year}
  {2004})}\BibitemShut {NoStop}%
\bibitem [{\citenamefont {Jarzynski}(2011)}]{Jarzynski2011}%
  \BibitemOpen
  \bibfield  {author} {\bibinfo {author} {\bibfnamefont {C.}~\bibnamefont
  {Jarzynski}},\ }\href {\doibase 10.1146/annurev-conmatphys-062910-140506}
  {\bibfield  {journal} {\bibinfo  {journal} {Annu. Rev. Condens. Matter
  Phys.}\ }\textbf {\bibinfo {volume} {2}},\ \bibinfo {pages} {329} (\bibinfo
  {year} {2011})}\BibitemShut {NoStop}%
\bibitem [{\citenamefont {Sekimoto}(2010)}]{Sekimoto2010}%
  \BibitemOpen
  \bibfield  {author} {\bibinfo {author} {\bibfnamefont {K.}~\bibnamefont
  {Sekimoto}},\ }\href {\doibase 10.1007/978-3-642-05411-2} {\emph {\bibinfo
  {title} {Stochastic Energetics}}}\ (\bibinfo  {publisher} {Springer Berlin
  Heidelberg},\ \bibinfo {year} {2010})\BibitemShut {NoStop}%
\bibitem [{\citenamefont {Seifert}(2012)}]{Seifert2012}%
  \BibitemOpen
  \bibfield  {author} {\bibinfo {author} {\bibfnamefont {U.}~\bibnamefont
  {Seifert}},\ }\href {\doibase 10.1088/0034-4885/75/12/126001} {\bibfield
  {journal} {\bibinfo  {journal} {Rep. Prog. Phys.}\ }\textbf {\bibinfo
  {volume} {75}},\ \bibinfo {pages} {126001} (\bibinfo {year}
  {2012})}\BibitemShut {NoStop}%
\bibitem [{\citenamefont {Seifert}(2005)}]{Seifert2005}%
  \BibitemOpen
  \bibfield  {author} {\bibinfo {author} {\bibfnamefont {U.}~\bibnamefont
  {Seifert}},\ }\href {\doibase 10.1103/physrevlett.95.040602} {\bibfield
  {journal} {\bibinfo  {journal} {Phys. Rev. Lett.}\ }\textbf {\bibinfo
  {volume} {95}},\ \bibinfo {pages} {040602} (\bibinfo {year}
  {2005})}\BibitemShut {NoStop}%
\bibitem [{\citenamefont {Esposito}\ \emph {et~al.}(2009)\citenamefont
  {Esposito}, \citenamefont {Harbola},\ and\ \citenamefont
  {Mukamel}}]{Esposito2009}%
  \BibitemOpen
  \bibfield  {author} {\bibinfo {author} {\bibfnamefont {M.}~\bibnamefont
  {Esposito}}, \bibinfo {author} {\bibfnamefont {U.}~\bibnamefont {Harbola}}, \
  and\ \bibinfo {author} {\bibfnamefont {S.}~\bibnamefont {Mukamel}},\ }\href
  {\doibase 10.1103/revmodphys.81.1665} {\bibfield  {journal} {\bibinfo
  {journal} {Rev. Mod. Phys.}\ }\textbf {\bibinfo {volume} {81}},\ \bibinfo
  {pages} {1665} (\bibinfo {year} {2009})}\BibitemShut {NoStop}%
\bibitem [{\citenamefont {Campisi}\ \emph {et~al.}(2011)\citenamefont
  {Campisi}, \citenamefont {H\"{a}nggi},\ and\ \citenamefont
  {Talkner}}]{Campisi2011}%
  \BibitemOpen
  \bibfield  {author} {\bibinfo {author} {\bibfnamefont {M.}~\bibnamefont
  {Campisi}}, \bibinfo {author} {\bibfnamefont {P.}~\bibnamefont {H\"{a}nggi}},
  \ and\ \bibinfo {author} {\bibfnamefont {P.}~\bibnamefont {Talkner}},\ }\href
  {\doibase 10.1103/revmodphys.83.771} {\bibfield  {journal} {\bibinfo
  {journal} {Rev. Mod. Phys.}\ }\textbf {\bibinfo {volume} {83}},\ \bibinfo
  {pages} {771} (\bibinfo {year} {2011})}\BibitemShut {NoStop}%
\bibitem [{\citenamefont {Klages}(2013)}]{Klages2013}%
  \BibitemOpen
  \bibfield  {author} {\bibinfo {author} {\bibfnamefont {R.}~\bibnamefont
  {Klages}},\ }\href@noop {} {\emph {\bibinfo {title} {Nonequilibrium
  statistical physics of small systems : fluctuation relations and beyond}}}\
  (\bibinfo  {publisher} {Wiley-VCH},\ \bibinfo {address} {Weinheim, Germany},\
  \bibinfo {year} {2013})\BibitemShut {NoStop}%
\bibitem [{\citenamefont {Horodecki}\ and\ \citenamefont
  {Oppenheim}(2013)}]{Horodecki2013}%
  \BibitemOpen
  \bibfield  {author} {\bibinfo {author} {\bibfnamefont {M.}~\bibnamefont
  {Horodecki}}\ and\ \bibinfo {author} {\bibfnamefont {J.}~\bibnamefont
  {Oppenheim}},\ }\href {\doibase 10.1038/ncomms3059} {\bibfield  {journal}
  {\bibinfo  {journal} {Nat Commun}\ }\textbf {\bibinfo {volume} {4}},\
  \bibinfo {pages} {2059} (\bibinfo {year} {2013})}\BibitemShut {NoStop}%
\bibitem [{\citenamefont {Ciliberto}(2017)}]{Ciliberto2017}%
  \BibitemOpen
  \bibfield  {author} {\bibinfo {author} {\bibfnamefont {S.}~\bibnamefont
  {Ciliberto}},\ }\href {\doibase 10.1103/physrevx.7.021051} {\bibfield
  {journal} {\bibinfo  {journal} {Phys. Rev. X}\ }\textbf {\bibinfo {volume}
  {7}},\ \bibinfo {pages} {021051} (\bibinfo {year} {2017})}\BibitemShut
  {NoStop}%
\bibitem [{\citenamefont {Tasaki}()}]{Tasaki2000}%
  \BibitemOpen
  \bibfield  {author} {\bibinfo {author} {\bibfnamefont {H.}~\bibnamefont
  {Tasaki}},\ }\href@noop {} {\ }\Eprint
  {http://arxiv.org/abs/cond-mat/0009244} {cond-mat/0009244} \BibitemShut
  {NoStop}%
\bibitem [{\citenamefont {Kurchan}(2000)}]{Kurchan2000}%
  \BibitemOpen
  \bibfield  {author} {\bibinfo {author} {\bibfnamefont {J.}~\bibnamefont
  {Kurchan}},\ }\href@noop {} {\  (\bibinfo {year} {2000})},\ \Eprint
  {http://arxiv.org/abs/cond-mat/0007360} {arXiv:cond-mat/0007360
  [cond-mat.stat-mech]} \BibitemShut {NoStop}%
\bibitem [{\citenamefont {Talkner}\ \emph {et~al.}(2007)\citenamefont
  {Talkner}, \citenamefont {Lutz},\ and\ \citenamefont
  {H\"{a}nggi}}]{Talkner2007}%
  \BibitemOpen
  \bibfield  {author} {\bibinfo {author} {\bibfnamefont {P.}~\bibnamefont
  {Talkner}}, \bibinfo {author} {\bibfnamefont {E.}~\bibnamefont {Lutz}}, \
  and\ \bibinfo {author} {\bibfnamefont {P.}~\bibnamefont {H\"{a}nggi}},\
  }\href {\doibase 10.1103/physreve.75.050102} {\bibfield  {journal} {\bibinfo
  {journal} {Phys. Rev. E}\ }\textbf {\bibinfo {volume} {75}},\ \bibinfo
  {pages} {050102} (\bibinfo {year} {2007})}\BibitemShut {NoStop}%
\bibitem [{\citenamefont {Deffner}\ and\ \citenamefont
  {Lutz}(2008)}]{Deffner2008}%
  \BibitemOpen
  \bibfield  {author} {\bibinfo {author} {\bibfnamefont {S.}~\bibnamefont
  {Deffner}}\ and\ \bibinfo {author} {\bibfnamefont {E.}~\bibnamefont {Lutz}},\
  }\href {\doibase 10.1103/physreve.77.021128} {\bibfield  {journal} {\bibinfo
  {journal} {Phys. Rev. E}\ }\textbf {\bibinfo {volume} {77}},\ \bibinfo
  {pages} {021128} (\bibinfo {year} {2008})}\BibitemShut {NoStop}%
\bibitem [{\citenamefont {Liu}(2014)}]{Liu2014}%
  \BibitemOpen
  \bibfield  {author} {\bibinfo {author} {\bibfnamefont {F.}~\bibnamefont
  {Liu}},\ }\href {\doibase 10.1103/physreve.90.032121} {\bibfield  {journal}
  {\bibinfo  {journal} {Phys. Rev. E}\ }\textbf {\bibinfo {volume} {90}},\
  \bibinfo {pages} {032121} (\bibinfo {year} {2014})}\BibitemShut {NoStop}%
\bibitem [{\citenamefont {Zhu}\ \emph {et~al.}(2016)\citenamefont {Zhu},
  \citenamefont {Gong}, \citenamefont {Wu},\ and\ \citenamefont
  {Quan}}]{Zhu2016}%
  \BibitemOpen
  \bibfield  {author} {\bibinfo {author} {\bibfnamefont {L.}~\bibnamefont
  {Zhu}}, \bibinfo {author} {\bibfnamefont {Z.}~\bibnamefont {Gong}}, \bibinfo
  {author} {\bibfnamefont {B.}~\bibnamefont {Wu}}, \ and\ \bibinfo {author}
  {\bibfnamefont {H.~T.}\ \bibnamefont {Quan}},\ }\href {\doibase
  10.1103/physreve.93.062108} {\bibfield  {journal} {\bibinfo  {journal} {Phys.
  Rev. E}\ }\textbf {\bibinfo {volume} {93}},\ \bibinfo {pages} {062108}
  (\bibinfo {year} {2016})}\BibitemShut {NoStop}%
\bibitem [{\citenamefont {Funo}\ and\ \citenamefont
  {Quan}(2018{\natexlab{a}})}]{Funo2018a}%
  \BibitemOpen
  \bibfield  {author} {\bibinfo {author} {\bibfnamefont {K.}~\bibnamefont
  {Funo}}\ and\ \bibinfo {author} {\bibfnamefont {H.}~\bibnamefont {Quan}},\
  }\href {\doibase 10.1103/physrevlett.121.040602} {\bibfield  {journal}
  {\bibinfo  {journal} {Phys. Rev. Lett.}\ }\textbf {\bibinfo {volume} {121}},\
  \bibinfo {pages} {040602} (\bibinfo {year} {2018}{\natexlab{a}})}\BibitemShut
  {NoStop}%
\bibitem [{\citenamefont {Salazar}\ and\ \citenamefont
  {Lira}(2019)}]{Salazar2019a}%
  \BibitemOpen
  \bibfield  {author} {\bibinfo {author} {\bibfnamefont {D.~S.~P.}\
  \bibnamefont {Salazar}}\ and\ \bibinfo {author} {\bibfnamefont {S.~A.}\
  \bibnamefont {Lira}},\ }\href {\doibase 10.1103/physreve.99.062119}
  {\bibfield  {journal} {\bibinfo  {journal} {Phys. Rev. E}\ }\textbf {\bibinfo
  {volume} {99}},\ \bibinfo {pages} {062119} (\bibinfo {year}
  {2019})}\BibitemShut {NoStop}%
\bibitem [{\citenamefont {Jarzynski}\ \emph {et~al.}(2015)\citenamefont
  {Jarzynski}, \citenamefont {Quan},\ and\ \citenamefont
  {Rahav}}]{Jarzynski2015}%
  \BibitemOpen
  \bibfield  {author} {\bibinfo {author} {\bibfnamefont {C.}~\bibnamefont
  {Jarzynski}}, \bibinfo {author} {\bibfnamefont {H.}~\bibnamefont {Quan}}, \
  and\ \bibinfo {author} {\bibfnamefont {S.}~\bibnamefont {Rahav}},\ }\href
  {\doibase 10.1103/physrevx.5.031038} {\bibfield  {journal} {\bibinfo
  {journal} {Phys. Rev. X}\ }\textbf {\bibinfo {volume} {5}},\ \bibinfo {pages}
  {031038} (\bibinfo {year} {2015})}\BibitemShut {NoStop}%
\bibitem [{\citenamefont {Deffner}\ \emph {et~al.}(2016)\citenamefont
  {Deffner}, \citenamefont {Paz},\ and\ \citenamefont {Zurek}}]{Deffner2016}%
  \BibitemOpen
  \bibfield  {author} {\bibinfo {author} {\bibfnamefont {S.}~\bibnamefont
  {Deffner}}, \bibinfo {author} {\bibfnamefont {J.~P.}\ \bibnamefont {Paz}}, \
  and\ \bibinfo {author} {\bibfnamefont {W.~H.}\ \bibnamefont {Zurek}},\ }\href
  {\doibase 10.1103/physreve.94.010103} {\bibfield  {journal} {\bibinfo
  {journal} {Phys. Rev. E}\ }\textbf {\bibinfo {volume} {94}},\ \bibinfo
  {pages} {010103(R)} (\bibinfo {year} {2016})}\BibitemShut {NoStop}%
\bibitem [{\citenamefont {Garc{\'{\i}}a-Mata}\ \emph
  {et~al.}(2017)\citenamefont {Garc{\'{\i}}a-Mata}, \citenamefont {Roncaglia},\
  and\ \citenamefont {Wisniacki}}]{GarciaMata2017}%
  \BibitemOpen
  \bibfield  {author} {\bibinfo {author} {\bibfnamefont {I.}~\bibnamefont
  {Garc{\'{\i}}a-Mata}}, \bibinfo {author} {\bibfnamefont {A.~J.}\ \bibnamefont
  {Roncaglia}}, \ and\ \bibinfo {author} {\bibfnamefont {D.~A.}\ \bibnamefont
  {Wisniacki}},\ }\href {\doibase 10.1103/physreve.95.050102} {\bibfield
  {journal} {\bibinfo  {journal} {Phys. Rev. E}\ }\textbf {\bibinfo {volume}
  {95}},\ \bibinfo {pages} {050102} (\bibinfo {year} {2017})}\BibitemShut
  {NoStop}%
\bibitem [{\citenamefont {Fei}\ and\ \citenamefont {Quan}(2020)}]{Fei2020}%
  \BibitemOpen
  \bibfield  {author} {\bibinfo {author} {\bibfnamefont {Z.}~\bibnamefont
  {Fei}}\ and\ \bibinfo {author} {\bibfnamefont {H.}~\bibnamefont {Quan}},\
  }\href {\doibase 10.1103/physrevlett.124.240603} {\bibfield  {journal}
  {\bibinfo  {journal} {Phys. Rev. Lett.}\ }\textbf {\bibinfo {volume} {124}},\
  \bibinfo {pages} {240603} (\bibinfo {year} {2020})}\BibitemShut {NoStop}%
\bibitem [{\citenamefont {Qiu}\ \emph {et~al.}(2020{\natexlab{a}})\citenamefont
  {Qiu}, \citenamefont {Fei}, \citenamefont {Pan},\ and\ \citenamefont
  {Quan}}]{Qiu2020}%
  \BibitemOpen
  \bibfield  {author} {\bibinfo {author} {\bibfnamefont {T.}~\bibnamefont
  {Qiu}}, \bibinfo {author} {\bibfnamefont {Z.}~\bibnamefont {Fei}}, \bibinfo
  {author} {\bibfnamefont {R.}~\bibnamefont {Pan}}, \ and\ \bibinfo {author}
  {\bibfnamefont {H.~T.}\ \bibnamefont {Quan}},\ }\href {\doibase
  10.1103/physreve.101.032111} {\bibfield  {journal} {\bibinfo  {journal}
  {Phys. Rev. E}\ }\textbf {\bibinfo {volume} {101}},\ \bibinfo {pages}
  {032111} (\bibinfo {year} {2020}{\natexlab{a}})}\BibitemShut {NoStop}%
\bibitem [{\citenamefont {Zurek}(1991)}]{Zurek1991}%
  \BibitemOpen
  \bibfield  {author} {\bibinfo {author} {\bibfnamefont {W.~H.}\ \bibnamefont
  {Zurek}},\ }\href {\doibase 10.1063/1.881293} {\bibfield  {journal} {\bibinfo
   {journal} {Phys. Today}\ }\textbf {\bibinfo {volume} {44}},\ \bibinfo
  {pages} {36} (\bibinfo {year} {1991})}\BibitemShut {NoStop}%
\bibitem [{\citenamefont {Zurek}(2003)}]{Zurek2003}%
  \BibitemOpen
  \bibfield  {author} {\bibinfo {author} {\bibfnamefont {W.~H.}\ \bibnamefont
  {Zurek}},\ }\href {\doibase 10.1103/revmodphys.75.715} {\bibfield  {journal}
  {\bibinfo  {journal} {Rev. Mod. Phys.}\ }\textbf {\bibinfo {volume} {75}},\
  \bibinfo {pages} {715} (\bibinfo {year} {2003})}\BibitemShut {NoStop}%
\bibitem [{\citenamefont {Saito}\ and\ \citenamefont {Dhar}(2007)}]{Saito2007}%
  \BibitemOpen
  \bibfield  {author} {\bibinfo {author} {\bibfnamefont {K.}~\bibnamefont
  {Saito}}\ and\ \bibinfo {author} {\bibfnamefont {A.}~\bibnamefont {Dhar}},\
  }\href {\doibase 10.1103/physrevlett.99.180601} {\bibfield  {journal}
  {\bibinfo  {journal} {Phys. Rev. Lett.}\ }\textbf {\bibinfo {volume} {99}},\
  \bibinfo {pages} {180601} (\bibinfo {year} {2007})}\BibitemShut {NoStop}%
\bibitem [{\citenamefont {Dubi}\ and\ \citenamefont {Ventra}(2011)}]{Dubi2011}%
  \BibitemOpen
  \bibfield  {author} {\bibinfo {author} {\bibfnamefont {Y.}~\bibnamefont
  {Dubi}}\ and\ \bibinfo {author} {\bibfnamefont {M.~D.}\ \bibnamefont
  {Ventra}},\ }\href {\doibase 10.1103/revmodphys.83.131} {\bibfield  {journal}
  {\bibinfo  {journal} {Rev. Mod. Phys.}\ }\textbf {\bibinfo {volume} {83}},\
  \bibinfo {pages} {131} (\bibinfo {year} {2011})}\BibitemShut {NoStop}%
\bibitem [{\citenamefont {Thingna}\ \emph {et~al.}(2012)\citenamefont
  {Thingna}, \citenamefont {Garc{\'{\i}}a-Palacios},\ and\ \citenamefont
  {Wang}}]{Thingna2012}%
  \BibitemOpen
  \bibfield  {author} {\bibinfo {author} {\bibfnamefont {J.}~\bibnamefont
  {Thingna}}, \bibinfo {author} {\bibfnamefont {J.~L.}\ \bibnamefont
  {Garc{\'{\i}}a-Palacios}}, \ and\ \bibinfo {author} {\bibfnamefont {J.-S.}\
  \bibnamefont {Wang}},\ }\href {\doibase 10.1103/physrevb.85.195452}
  {\bibfield  {journal} {\bibinfo  {journal} {Phys. Rev. B}\ }\textbf {\bibinfo
  {volume} {85}},\ \bibinfo {pages} {195452} (\bibinfo {year}
  {2012})}\BibitemShut {NoStop}%
\bibitem [{\citenamefont {Wang}\ \emph {et~al.}(2014)\citenamefont {Wang},
  \citenamefont {Agarwalla}, \citenamefont {Li},\ and\ \citenamefont
  {Thingna}}]{Wang2013}%
  \BibitemOpen
  \bibfield  {author} {\bibinfo {author} {\bibfnamefont {J.-S.}\ \bibnamefont
  {Wang}}, \bibinfo {author} {\bibfnamefont {B.~K.}\ \bibnamefont {Agarwalla}},
  \bibinfo {author} {\bibfnamefont {H.}~\bibnamefont {Li}}, \ and\ \bibinfo
  {author} {\bibfnamefont {J.}~\bibnamefont {Thingna}},\ }\href {\doibase
  10.1007/s11467-013-0340-x} {\bibfield  {journal} {\bibinfo  {journal} {Front.
  Phys.}\ }\textbf {\bibinfo {volume} {9}},\ \bibinfo {pages} {673} (\bibinfo
  {year} {2014})}\BibitemShut {NoStop}%
\bibitem [{\citenamefont {Thingna}\ \emph {et~al.}(2016)\citenamefont
  {Thingna}, \citenamefont {Manzano},\ and\ \citenamefont {Cao}}]{Thingna2016}%
  \BibitemOpen
  \bibfield  {author} {\bibinfo {author} {\bibfnamefont {J.}~\bibnamefont
  {Thingna}}, \bibinfo {author} {\bibfnamefont {D.}~\bibnamefont {Manzano}}, \
  and\ \bibinfo {author} {\bibfnamefont {J.}~\bibnamefont {Cao}},\ }\href
  {\doibase 10.1038/srep28027} {\bibfield  {journal} {\bibinfo  {journal} {Sci.
  Rep.}\ }\textbf {\bibinfo {volume} {6}},\ \bibinfo {pages} {28027} (\bibinfo
  {year} {2016})}\BibitemShut {NoStop}%
\bibitem [{\citenamefont {He}\ \emph {et~al.}(2016)\citenamefont {He},
  \citenamefont {Thingna}, \citenamefont {Wang},\ and\ \citenamefont
  {Li}}]{He2016}%
  \BibitemOpen
  \bibfield  {author} {\bibinfo {author} {\bibfnamefont {D.}~\bibnamefont
  {He}}, \bibinfo {author} {\bibfnamefont {J.}~\bibnamefont {Thingna}},
  \bibinfo {author} {\bibfnamefont {J.-S.}\ \bibnamefont {Wang}}, \ and\
  \bibinfo {author} {\bibfnamefont {B.}~\bibnamefont {Li}},\ }\href {\doibase
  10.1103/physrevb.94.155411} {\bibfield  {journal} {\bibinfo  {journal} {Phys.
  Rev. B}\ }\textbf {\bibinfo {volume} {94}},\ \bibinfo {pages} {155411}
  (\bibinfo {year} {2016})}\BibitemShut {NoStop}%
\bibitem [{\citenamefont {Segal}\ and\ \citenamefont
  {Agarwalla}(2016)}]{Segal2016}%
  \BibitemOpen
  \bibfield  {author} {\bibinfo {author} {\bibfnamefont {D.}~\bibnamefont
  {Segal}}\ and\ \bibinfo {author} {\bibfnamefont {B.~K.}\ \bibnamefont
  {Agarwalla}},\ }\href {\doibase 10.1146/annurev-physchem-040215-112103}
  {\bibfield  {journal} {\bibinfo  {journal} {Ann. Phys. Chem.}\ }\textbf
  {\bibinfo {volume} {67}},\ \bibinfo {pages} {185} (\bibinfo {year}
  {2016})}\BibitemShut {NoStop}%
\bibitem [{\citenamefont {Kilgour}\ \emph {et~al.}(2019)\citenamefont
  {Kilgour}, \citenamefont {Agarwalla},\ and\ \citenamefont
  {Segal}}]{Kilgour2019}%
  \BibitemOpen
  \bibfield  {author} {\bibinfo {author} {\bibfnamefont {M.}~\bibnamefont
  {Kilgour}}, \bibinfo {author} {\bibfnamefont {B.~K.}\ \bibnamefont
  {Agarwalla}}, \ and\ \bibinfo {author} {\bibfnamefont {D.}~\bibnamefont
  {Segal}},\ }\href {\doibase 10.1063/1.5084949} {\bibfield  {journal}
  {\bibinfo  {journal} {J. Chem. Phys.}\ }\textbf {\bibinfo {volume} {150}},\
  \bibinfo {pages} {084111} (\bibinfo {year} {2019})}\BibitemShut {NoStop}%
\bibitem [{\citenamefont {Wang}\ \emph {et~al.}(2017)\citenamefont {Wang},
  \citenamefont {Ren},\ and\ \citenamefont {Cao}}]{Wang2017}%
  \BibitemOpen
  \bibfield  {author} {\bibinfo {author} {\bibfnamefont {C.}~\bibnamefont
  {Wang}}, \bibinfo {author} {\bibfnamefont {J.}~\bibnamefont {Ren}}, \ and\
  \bibinfo {author} {\bibfnamefont {J.}~\bibnamefont {Cao}},\ }\href {\doibase
  10.1103/physreva.95.023610} {\bibfield  {journal} {\bibinfo  {journal} {Phys.
  Rev. A}\ }\textbf {\bibinfo {volume} {95}},\ \bibinfo {pages} {023610}
  (\bibinfo {year} {2017})}\BibitemShut {NoStop}%
\bibitem [{\citenamefont {Aurell}\ \emph {et~al.}(2020)\citenamefont {Aurell},
  \citenamefont {Donvil},\ and\ \citenamefont {Mallick}}]{Aurell2020}%
  \BibitemOpen
  \bibfield  {author} {\bibinfo {author} {\bibfnamefont {E.}~\bibnamefont
  {Aurell}}, \bibinfo {author} {\bibfnamefont {B.}~\bibnamefont {Donvil}}, \
  and\ \bibinfo {author} {\bibfnamefont {K.}~\bibnamefont {Mallick}},\ }\href
  {\doibase 10.1103/physreve.101.052116} {\bibfield  {journal} {\bibinfo
  {journal} {Phys. Rev. E}\ }\textbf {\bibinfo {volume} {101}},\ \bibinfo
  {pages} {052116} (\bibinfo {year} {2020})}\BibitemShut {NoStop}%
\bibitem [{\citenamefont {Denzler}\ and\ \citenamefont
  {Lutz}(2018)}]{Denzler2018}%
  \BibitemOpen
  \bibfield  {author} {\bibinfo {author} {\bibfnamefont {T.}~\bibnamefont
  {Denzler}}\ and\ \bibinfo {author} {\bibfnamefont {E.}~\bibnamefont {Lutz}},\
  }\href {\doibase 10.1103/physreve.98.052106} {\bibfield  {journal} {\bibinfo
  {journal} {Phys. Rev. E}\ }\textbf {\bibinfo {volume} {98}},\ \bibinfo
  {pages} {052106} (\bibinfo {year} {2018})}\BibitemShut {NoStop}%
\bibitem [{\citenamefont {Salazar}\ \emph {et~al.}(2019)\citenamefont
  {Salazar}, \citenamefont {Mac{\^{e}}do},\ and\ \citenamefont
  {Vasconcelos}}]{Salazar2019}%
  \BibitemOpen
  \bibfield  {author} {\bibinfo {author} {\bibfnamefont {D.~S.~P.}\
  \bibnamefont {Salazar}}, \bibinfo {author} {\bibfnamefont {A.~M.~S.}\
  \bibnamefont {Mac{\^{e}}do}}, \ and\ \bibinfo {author} {\bibfnamefont
  {G.~L.}\ \bibnamefont {Vasconcelos}},\ }\href {\doibase
  10.1103/physreve.99.022133} {\bibfield  {journal} {\bibinfo  {journal} {Phys.
  Rev. E}\ }\textbf {\bibinfo {volume} {99}},\ \bibinfo {pages} {022133}
  (\bibinfo {year} {2019})}\BibitemShut {NoStop}%
\bibitem [{\citenamefont {Popovic}\ \emph {et~al.}(2021)\citenamefont
  {Popovic}, \citenamefont {Mitchison}, \citenamefont {Strathearn},
  \citenamefont {Lovett}, \citenamefont {Goold},\ and\ \citenamefont
  {Eastham}}]{Popovic2021}%
  \BibitemOpen
  \bibfield  {author} {\bibinfo {author} {\bibfnamefont {M.}~\bibnamefont
  {Popovic}}, \bibinfo {author} {\bibfnamefont {M.~T.}\ \bibnamefont
  {Mitchison}}, \bibinfo {author} {\bibfnamefont {A.}~\bibnamefont
  {Strathearn}}, \bibinfo {author} {\bibfnamefont {B.~W.}\ \bibnamefont
  {Lovett}}, \bibinfo {author} {\bibfnamefont {J.}~\bibnamefont {Goold}}, \
  and\ \bibinfo {author} {\bibfnamefont {P.~R.}\ \bibnamefont {Eastham}},\
  }\href {\doibase 10.1103/prxquantum.2.020338} {\bibfield  {journal} {\bibinfo
   {journal} {PRX Quantum}\ }\textbf {\bibinfo {volume} {2}},\ \bibinfo {pages}
  {020338} (\bibinfo {year} {2021})}\BibitemShut {NoStop}%
\bibitem [{\citenamefont {Karsten~Balzer}(2012)}]{KarstenBalzer2012}%
  \BibitemOpen
  \bibfield  {author} {\bibinfo {author} {\bibfnamefont {M.~B.}\ \bibnamefont
  {Karsten~Balzer}},\ }\href
  {https://www.ebook.de/de/product/19841336/karsten_balzer_michael_bonitz_nonequilibrium_green_s_functions_approach_to_inhomogeneous_systems.html}
  {\emph {\bibinfo {title} {Nonequilibrium Green's Functions Approach to
  Inhomogeneous Systems}}}\ (\bibinfo  {publisher} {Springer Berlin
  Heidelberg},\ \bibinfo {year} {2012})\BibitemShut {NoStop}%
\bibitem [{\citenamefont {Esposito}\ \emph
  {et~al.}(2015{\natexlab{a}})\citenamefont {Esposito}, \citenamefont {Ochoa},\
  and\ \citenamefont {Galperin}}]{Esposito2015a}%
  \BibitemOpen
  \bibfield  {author} {\bibinfo {author} {\bibfnamefont {M.}~\bibnamefont
  {Esposito}}, \bibinfo {author} {\bibfnamefont {M.~A.}\ \bibnamefont {Ochoa}},
  \ and\ \bibinfo {author} {\bibfnamefont {M.}~\bibnamefont {Galperin}},\
  }\href {\doibase 10.1103/physrevlett.114.080602} {\bibfield  {journal}
  {\bibinfo  {journal} {Phys. Rev. Lett.}\ }\textbf {\bibinfo {volume} {114}},\
  \bibinfo {pages} {080602} (\bibinfo {year} {2015}{\natexlab{a}})}\BibitemShut
  {NoStop}%
\bibitem [{\citenamefont {Polanco}(2021)}]{Polanco2021}%
  \BibitemOpen
  \bibfield  {author} {\bibinfo {author} {\bibfnamefont {C.~A.}\ \bibnamefont
  {Polanco}},\ }\href {\doibase 10.1080/15567265.2021.1881193} {\bibfield
  {journal} {\bibinfo  {journal} {Nanoscale Microscale Thermophys. Eng.}\
  }\textbf {\bibinfo {volume} {25}},\ \bibinfo {pages} {1} (\bibinfo {year}
  {2021})}\BibitemShut {NoStop}%
\bibitem [{\citenamefont {Aron}\ \emph {et~al.}(2010)\citenamefont {Aron},
  \citenamefont {Biroli},\ and\ \citenamefont {Cugliandolo}}]{Aron2010}%
  \BibitemOpen
  \bibfield  {author} {\bibinfo {author} {\bibfnamefont {C.}~\bibnamefont
  {Aron}}, \bibinfo {author} {\bibfnamefont {G.}~\bibnamefont {Biroli}}, \ and\
  \bibinfo {author} {\bibfnamefont {L.~F.}\ \bibnamefont {Cugliandolo}},\
  }\href {\doibase 10.1088/1742-5468/2010/11/p11018} {\bibfield  {journal}
  {\bibinfo  {journal} {J. Stat. Mech.}\ }\textbf {\bibinfo {volume} {2010}},\
  \bibinfo {pages} {P11018} (\bibinfo {year} {2010})}\BibitemShut {NoStop}%
\bibitem [{\citenamefont {Mallick}\ \emph {et~al.}(2011)\citenamefont
  {Mallick}, \citenamefont {Moshe},\ and\ \citenamefont
  {Orland}}]{Mallick2011}%
  \BibitemOpen
  \bibfield  {author} {\bibinfo {author} {\bibfnamefont {K.}~\bibnamefont
  {Mallick}}, \bibinfo {author} {\bibfnamefont {M.}~\bibnamefont {Moshe}}, \
  and\ \bibinfo {author} {\bibfnamefont {H.}~\bibnamefont {Orland}},\ }\href
  {\doibase 10.1088/1751-8113/44/9/095002} {\bibfield  {journal} {\bibinfo
  {journal} {J. Phys. A}\ }\textbf {\bibinfo {volume} {44}},\ \bibinfo {pages}
  {095002} (\bibinfo {year} {2011})}\BibitemShut {NoStop}%
\bibitem [{\citenamefont {Carrega}\ \emph {et~al.}(2015)\citenamefont
  {Carrega}, \citenamefont {Solinas}, \citenamefont {Braggio}, \citenamefont
  {Sassetti},\ and\ \citenamefont {Weiss}}]{Carrega2015}%
  \BibitemOpen
  \bibfield  {author} {\bibinfo {author} {\bibfnamefont {M.}~\bibnamefont
  {Carrega}}, \bibinfo {author} {\bibfnamefont {P.}~\bibnamefont {Solinas}},
  \bibinfo {author} {\bibfnamefont {A.}~\bibnamefont {Braggio}}, \bibinfo
  {author} {\bibfnamefont {M.}~\bibnamefont {Sassetti}}, \ and\ \bibinfo
  {author} {\bibfnamefont {U.}~\bibnamefont {Weiss}},\ }\href {\doibase
  10.1088/1367-2630/17/4/045030} {\bibfield  {journal} {\bibinfo  {journal}
  {New J. Phys.}\ }\textbf {\bibinfo {volume} {17}},\ \bibinfo {pages} {045030}
  (\bibinfo {year} {2015})}\BibitemShut {NoStop}%
\bibitem [{\citenamefont {Funo}\ and\ \citenamefont
  {Quan}(2018{\natexlab{b}})}]{Funo2018}%
  \BibitemOpen
  \bibfield  {author} {\bibinfo {author} {\bibfnamefont {K.}~\bibnamefont
  {Funo}}\ and\ \bibinfo {author} {\bibfnamefont {H.~T.}\ \bibnamefont
  {Quan}},\ }\href {\doibase 10.1103/physreve.98.012113} {\bibfield  {journal}
  {\bibinfo  {journal} {Phys. Rev. E}\ }\textbf {\bibinfo {volume} {98}},\
  \bibinfo {pages} {012113} (\bibinfo {year} {2018}{\natexlab{b}})}\BibitemShut
  {NoStop}%
\bibitem [{\citenamefont {Yeo}(2019)}]{Yeo2019}%
  \BibitemOpen
  \bibfield  {author} {\bibinfo {author} {\bibfnamefont {J.}~\bibnamefont
  {Yeo}},\ }\href {\doibase 10.1103/physreve.100.062107} {\bibfield  {journal}
  {\bibinfo  {journal} {Phys. Rev. E}\ }\textbf {\bibinfo {volume} {100}},\
  \bibinfo {pages} {062107} (\bibinfo {year} {2019})}\BibitemShut {NoStop}%
\bibitem [{\citenamefont {Fogedby}(2020)}]{Fogedby2020}%
  \BibitemOpen
  \bibfield  {author} {\bibinfo {author} {\bibfnamefont {H.~C.}\ \bibnamefont
  {Fogedby}},\ }\href {\doibase 10.1088/1742-5468/aba7b2} {\bibfield  {journal}
  {\bibinfo  {journal} {J. Stat. Mech.: Theory Exp.}\ }\textbf {\bibinfo
  {volume} {2020}},\ \bibinfo {pages} {083208} (\bibinfo {year}
  {2020})}\BibitemShut {NoStop}%
\bibitem [{\citenamefont {van Zon}\ and\ \citenamefont
  {Cohen}(2004)}]{Zon2004}%
  \BibitemOpen
  \bibfield  {author} {\bibinfo {author} {\bibfnamefont {R.}~\bibnamefont {van
  Zon}}\ and\ \bibinfo {author} {\bibfnamefont {E.~G.~D.}\ \bibnamefont
  {Cohen}},\ }\href {\doibase 10.1103/physreve.69.056121} {\bibfield  {journal}
  {\bibinfo  {journal} {Phys. Rev. E}\ }\textbf {\bibinfo {volume} {69}},\
  \bibinfo {pages} {056121} (\bibinfo {year} {2004})}\BibitemShut {NoStop}%
\bibitem [{\citenamefont {Imparato}\ \emph {et~al.}(2007)\citenamefont
  {Imparato}, \citenamefont {Peliti}, \citenamefont {Pesce}, \citenamefont
  {Rusciano},\ and\ \citenamefont {Sasso}}]{Imparato2007}%
  \BibitemOpen
  \bibfield  {author} {\bibinfo {author} {\bibfnamefont {A.}~\bibnamefont
  {Imparato}}, \bibinfo {author} {\bibfnamefont {L.}~\bibnamefont {Peliti}},
  \bibinfo {author} {\bibfnamefont {G.}~\bibnamefont {Pesce}}, \bibinfo
  {author} {\bibfnamefont {G.}~\bibnamefont {Rusciano}}, \ and\ \bibinfo
  {author} {\bibfnamefont {A.}~\bibnamefont {Sasso}},\ }\href {\doibase
  10.1103/physreve.76.050101} {\bibfield  {journal} {\bibinfo  {journal} {Phys.
  Rev. E}\ }\textbf {\bibinfo {volume} {76}},\ \bibinfo {pages} {050101}
  (\bibinfo {year} {2007})}\BibitemShut {NoStop}%
\bibitem [{\citenamefont {Fogedby}\ and\ \citenamefont
  {Imparato}(2009)}]{Fogedby2009}%
  \BibitemOpen
  \bibfield  {author} {\bibinfo {author} {\bibfnamefont {H.~C.}\ \bibnamefont
  {Fogedby}}\ and\ \bibinfo {author} {\bibfnamefont {A.}~\bibnamefont
  {Imparato}},\ }\href {\doibase 10.1088/1751-8113/42/47/475004} {\bibfield
  {journal} {\bibinfo  {journal} {J. Phys. A: Math. Theor.}\ }\textbf {\bibinfo
  {volume} {42}},\ \bibinfo {pages} {475004} (\bibinfo {year}
  {2009})}\BibitemShut {NoStop}%
\bibitem [{\citenamefont {Chatterjee}\ and\ \citenamefont
  {Cherayil}(2010)}]{Chatterjee2010}%
  \BibitemOpen
  \bibfield  {author} {\bibinfo {author} {\bibfnamefont {D.}~\bibnamefont
  {Chatterjee}}\ and\ \bibinfo {author} {\bibfnamefont {B.~J.}\ \bibnamefont
  {Cherayil}},\ }\href {\doibase 10.1103/physreve.82.051104} {\bibfield
  {journal} {\bibinfo  {journal} {Phys. Rev. E}\ }\textbf {\bibinfo {volume}
  {82}},\ \bibinfo {pages} {051104} (\bibinfo {year} {2010})}\BibitemShut
  {NoStop}%
\bibitem [{\citenamefont {Gomez-Solano}\ \emph {et~al.}(2011)\citenamefont
  {Gomez-Solano}, \citenamefont {Petrosyan},\ and\ \citenamefont
  {Ciliberto}}]{GomezSolano2011}%
  \BibitemOpen
  \bibfield  {author} {\bibinfo {author} {\bibfnamefont {J.~R.}\ \bibnamefont
  {Gomez-Solano}}, \bibinfo {author} {\bibfnamefont {A.}~\bibnamefont
  {Petrosyan}}, \ and\ \bibinfo {author} {\bibfnamefont {S.}~\bibnamefont
  {Ciliberto}},\ }\href {\doibase 10.1103/physrevlett.106.200602} {\bibfield
  {journal} {\bibinfo  {journal} {Phys. Rev. Lett.}\ }\textbf {\bibinfo
  {volume} {106}},\ \bibinfo {pages} {200602} (\bibinfo {year}
  {2011})}\BibitemShut {NoStop}%
\bibitem [{\citenamefont {Salazar}\ and\ \citenamefont
  {Lira}(2016)}]{Salazar2016}%
  \BibitemOpen
  \bibfield  {author} {\bibinfo {author} {\bibfnamefont {D.~S.~P.}\
  \bibnamefont {Salazar}}\ and\ \bibinfo {author} {\bibfnamefont {S.~A.}\
  \bibnamefont {Lira}},\ }\href {\doibase 10.1088/1751-8113/49/46/465001}
  {\bibfield  {journal} {\bibinfo  {journal} {J. Phys. A: Math. Theor.}\
  }\textbf {\bibinfo {volume} {49}},\ \bibinfo {pages} {465001} (\bibinfo
  {year} {2016})}\BibitemShut {NoStop}%
\bibitem [{\citenamefont {Pagare}\ and\ \citenamefont
  {Cherayil}(2019)}]{Pagare2019}%
  \BibitemOpen
  \bibfield  {author} {\bibinfo {author} {\bibfnamefont {A.}~\bibnamefont
  {Pagare}}\ and\ \bibinfo {author} {\bibfnamefont {B.~J.}\ \bibnamefont
  {Cherayil}},\ }\href {\doibase 10.1103/physreve.100.052124} {\bibfield
  {journal} {\bibinfo  {journal} {Phys. Rev. E}\ }\textbf {\bibinfo {volume}
  {100}},\ \bibinfo {pages} {052124} (\bibinfo {year} {2019})}\BibitemShut
  {NoStop}%
\bibitem [{\citenamefont {Paraguass{\'{u}}}\ \emph {et~al.}(2021)\citenamefont
  {Paraguass{\'{u}}}, \citenamefont {Aquino},\ and\ \citenamefont
  {Morgado}}]{Paraguassu2021a}%
  \BibitemOpen
  \bibfield  {author} {\bibinfo {author} {\bibfnamefont {P.~V.}\ \bibnamefont
  {Paraguass{\'{u}}}}, \bibinfo {author} {\bibfnamefont {R.}~\bibnamefont
  {Aquino}}, \ and\ \bibinfo {author} {\bibfnamefont {W.~A.~M.}\ \bibnamefont
  {Morgado}},\ }\href@noop {} {\  (\bibinfo {year} {2021})},\ \Eprint
  {http://arxiv.org/abs/2102.09115} {arXiv:2102.09115 [cond-mat.stat-mech]}
  \BibitemShut {NoStop}%
\bibitem [{\citenamefont {Gupta}\ and\ \citenamefont
  {Sivak}(2021)}]{Gupta2021}%
  \BibitemOpen
  \bibfield  {author} {\bibinfo {author} {\bibfnamefont {D.}~\bibnamefont
  {Gupta}}\ and\ \bibinfo {author} {\bibfnamefont {D.~A.}\ \bibnamefont
  {Sivak}},\ }\href@noop {} {\  (\bibinfo {year} {2021})},\ \Eprint
  {http://arxiv.org/abs/2103.09358} {arXiv:2103.09358 [cond-mat.stat-mech]}
  \BibitemShut {NoStop}%
\bibitem [{\citenamefont {Esposito}\ \emph
  {et~al.}(2015{\natexlab{b}})\citenamefont {Esposito}, \citenamefont {Ochoa},\
  and\ \citenamefont {Galperin}}]{Esposito2015}%
  \BibitemOpen
  \bibfield  {author} {\bibinfo {author} {\bibfnamefont {M.}~\bibnamefont
  {Esposito}}, \bibinfo {author} {\bibfnamefont {M.~A.}\ \bibnamefont {Ochoa}},
  \ and\ \bibinfo {author} {\bibfnamefont {M.}~\bibnamefont {Galperin}},\
  }\href {\doibase 10.1103/physrevb.92.235440} {\bibfield  {journal} {\bibinfo
  {journal} {Phys. Rev. B}\ }\textbf {\bibinfo {volume} {92}},\ \bibinfo
  {pages} {235440} (\bibinfo {year} {2015}{\natexlab{b}})}\BibitemShut
  {NoStop}%
\bibitem [{\citenamefont {Talkner}\ and\ \citenamefont
  {H\"{a}nggi}(2016)}]{Talkner2016a}%
  \BibitemOpen
  \bibfield  {author} {\bibinfo {author} {\bibfnamefont {P.}~\bibnamefont
  {Talkner}}\ and\ \bibinfo {author} {\bibfnamefont {P.}~\bibnamefont
  {H\"{a}nggi}},\ }\href {\doibase 10.1103/physreve.94.022143} {\bibfield
  {journal} {\bibinfo  {journal} {Phys. Rev. E}\ }\textbf {\bibinfo {volume}
  {94}},\ \bibinfo {pages} {022143} (\bibinfo {year} {2016})}\BibitemShut
  {NoStop}%
\bibitem [{\citenamefont {Talkner}\ and\ \citenamefont
  {H\"{a}nggi}(2020)}]{Talkner2020}%
  \BibitemOpen
  \bibfield  {author} {\bibinfo {author} {\bibfnamefont {P.}~\bibnamefont
  {Talkner}}\ and\ \bibinfo {author} {\bibfnamefont {P.}~\bibnamefont
  {H\"{a}nggi}},\ }\href {\doibase 10.1103/revmodphys.92.041002} {\bibfield
  {journal} {\bibinfo  {journal} {Rev. Mod. Phys.}\ }\textbf {\bibinfo {volume}
  {92}},\ \bibinfo {pages} {041002} (\bibinfo {year} {2020})}\BibitemShut
  {NoStop}%
\bibitem [{\citenamefont {Bez}(1980)}]{Bez1980}%
  \BibitemOpen
  \bibfield  {author} {\bibinfo {author} {\bibfnamefont {W.}~\bibnamefont
  {Bez}},\ }\href {\doibase 10.1007/bf01305831} {\bibfield  {journal} {\bibinfo
   {journal} {Z. Phys. B}\ }\textbf {\bibinfo {volume} {39}},\ \bibinfo {pages}
  {319} (\bibinfo {year} {1980})}\BibitemShut {NoStop}%
\bibitem [{\citenamefont {Caldeira}\ and\ \citenamefont
  {Leggett}(1983{\natexlab{a}})}]{Caldeira1983}%
  \BibitemOpen
  \bibfield  {author} {\bibinfo {author} {\bibfnamefont {A.}~\bibnamefont
  {Caldeira}}\ and\ \bibinfo {author} {\bibfnamefont {A.}~\bibnamefont
  {Leggett}},\ }\href {\doibase 10.1016/0378-4371(83)90013-4} {\bibfield
  {journal} {\bibinfo  {journal} {Phys. A}\ }\textbf {\bibinfo {volume}
  {121}},\ \bibinfo {pages} {587} (\bibinfo {year}
  {1983}{\natexlab{a}})}\BibitemShut {NoStop}%
\bibitem [{\citenamefont {Caldeira}\ and\ \citenamefont
  {Leggett}(1983{\natexlab{b}})}]{Caldeira1983b}%
  \BibitemOpen
  \bibfield  {author} {\bibinfo {author} {\bibfnamefont {A.}~\bibnamefont
  {Caldeira}}\ and\ \bibinfo {author} {\bibfnamefont {A.}~\bibnamefont
  {Leggett}},\ }\href {\doibase 10.1016/0003-4916(83)90202-6} {\bibfield
  {journal} {\bibinfo  {journal} {Ann. Phys.}\ }\textbf {\bibinfo {volume}
  {149}},\ \bibinfo {pages} {374} (\bibinfo {year}
  {1983}{\natexlab{b}})}\BibitemShut {NoStop}%
\bibitem [{\citenamefont {Unruh}\ and\ \citenamefont
  {Zurek}(1989)}]{Unruh1989}%
  \BibitemOpen
  \bibfield  {author} {\bibinfo {author} {\bibfnamefont {W.~G.}\ \bibnamefont
  {Unruh}}\ and\ \bibinfo {author} {\bibfnamefont {W.~H.}\ \bibnamefont
  {Zurek}},\ }\href {\doibase 10.1103/physrevd.40.1071} {\bibfield  {journal}
  {\bibinfo  {journal} {Phys. Rev. D}\ }\textbf {\bibinfo {volume} {40}},\
  \bibinfo {pages} {1071} (\bibinfo {year} {1989})}\BibitemShut {NoStop}%
\bibitem [{\citenamefont {Breuer}\ and\ \citenamefont
  {Petruccione}(2007)}]{Breuer2007}%
  \BibitemOpen
  \bibfield  {author} {\bibinfo {author} {\bibfnamefont {H.-P.}\ \bibnamefont
  {Breuer}}\ and\ \bibinfo {author} {\bibfnamefont {F.}~\bibnamefont
  {Petruccione}},\ }\href {\doibase 10.1093/acprof:oso/9780199213900.001.0001}
  {\emph {\bibinfo {title} {The Theory of Open Quantum Systems}}}\ (\bibinfo
  {publisher} {Oxford University Press},\ \bibinfo {year} {2007})\BibitemShut
  {NoStop}%
\bibitem [{\citenamefont {Weiss}(2008)}]{Weiss2008}%
  \BibitemOpen
  \bibfield  {author} {\bibinfo {author} {\bibfnamefont {U.}~\bibnamefont
  {Weiss}},\ }\href
  {https://www.ebook.de/de/product/7191616/ulrich_weiss_quantum_dissipative_systems.html}
  {\emph {\bibinfo {title} {Quantum Dissipative Systems}}}\ (\bibinfo
  {publisher} {World Scientific Publishing Company},\ \bibinfo {year}
  {2008})\BibitemShut {NoStop}%
\bibitem [{Note1()}]{Note1}%
  \BibitemOpen
  \bibinfo {note} {Usually the quantum fluctuating heat is defined via
  two-point measurements over the heat bath. When the Hamiltonian of the system
  is time-independent, the internal energy change of the system is completely
  caused by the heat exchange. The quantum fluctuating heat can thus be
  alternatively defined via two-point measurements over the system, whose
  number of degrees of freedom is much smaller than that of the heat bath.
  Hence, the calculation of the heat statistics can be significantly simplified
  under this definition.}\BibitemShut {Stop}%
\bibitem [{\citenamefont {Yu}\ and\ \citenamefont {Sun}(1994)}]{Yu1994}%
  \BibitemOpen
  \bibfield  {author} {\bibinfo {author} {\bibfnamefont {L.~H.}\ \bibnamefont
  {Yu}}\ and\ \bibinfo {author} {\bibfnamefont {C.-P.}\ \bibnamefont {Sun}},\
  }\href {\doibase 10.1103/physreva.49.592} {\bibfield  {journal} {\bibinfo
  {journal} {Phys. Rev. A}\ }\textbf {\bibinfo {volume} {49}},\ \bibinfo
  {pages} {592} (\bibinfo {year} {1994})}\BibitemShut {NoStop}%
\bibitem [{\citenamefont {Wigner}(1932)}]{Wigner1932}%
  \BibitemOpen
  \bibfield  {author} {\bibinfo {author} {\bibfnamefont {E.}~\bibnamefont
  {Wigner}},\ }\href {\doibase 10.1103/physrev.40.749} {\bibfield  {journal}
  {\bibinfo  {journal} {Phys. Rev.}\ }\textbf {\bibinfo {volume} {40}},\
  \bibinfo {pages} {749} (\bibinfo {year} {1932})}\BibitemShut {NoStop}%
\bibitem [{\citenamefont {Hillery}\ \emph {et~al.}(1984)\citenamefont
  {Hillery}, \citenamefont {O'Connell}, \citenamefont {Scully},\ and\
  \citenamefont {Wigner}}]{Hillery_1984}%
  \BibitemOpen
  \bibfield  {author} {\bibinfo {author} {\bibfnamefont {M.}~\bibnamefont
  {Hillery}}, \bibinfo {author} {\bibfnamefont {R.}~\bibnamefont {O'Connell}},
  \bibinfo {author} {\bibfnamefont {M.}~\bibnamefont {Scully}}, \ and\ \bibinfo
  {author} {\bibfnamefont {E.}~\bibnamefont {Wigner}},\ }\href {\doibase
  10.1016/0370-1573(84)90160-1} {\bibfield  {journal} {\bibinfo  {journal}
  {Phys. Rep.}\ }\textbf {\bibinfo {volume} {106}},\ \bibinfo {pages} {121}
  (\bibinfo {year} {1984})}\BibitemShut {NoStop}%
\bibitem [{\citenamefont {Polkovnikov}(2010)}]{Polkovnikov2010}%
  \BibitemOpen
  \bibfield  {author} {\bibinfo {author} {\bibfnamefont {A.}~\bibnamefont
  {Polkovnikov}},\ }\href {\doibase 10.1016/j.aop.2010.02.006} {\bibfield
  {journal} {\bibinfo  {journal} {Ann. Phys.}\ }\textbf {\bibinfo {volume}
  {325}},\ \bibinfo {pages} {1790} (\bibinfo {year} {2010})}\BibitemShut
  {NoStop}%
\bibitem [{\citenamefont {Fei}\ \emph {et~al.}(2018)\citenamefont {Fei},
  \citenamefont {Quan},\ and\ \citenamefont {Liu}}]{Fei2018}%
  \BibitemOpen
  \bibfield  {author} {\bibinfo {author} {\bibfnamefont {Z.}~\bibnamefont
  {Fei}}, \bibinfo {author} {\bibfnamefont {H.~T.}\ \bibnamefont {Quan}}, \
  and\ \bibinfo {author} {\bibfnamefont {F.}~\bibnamefont {Liu}},\ }\href
  {\doibase 10.1103/physreve.98.012132} {\bibfield  {journal} {\bibinfo
  {journal} {Phys. Rev. E}\ }\textbf {\bibinfo {volume} {98}},\ \bibinfo
  {pages} {012132} (\bibinfo {year} {2018})}\BibitemShut {NoStop}%
\bibitem [{\citenamefont {Qian}\ and\ \citenamefont {Liu}(2019)}]{Qian2019}%
  \BibitemOpen
  \bibfield  {author} {\bibinfo {author} {\bibfnamefont {Y.}~\bibnamefont
  {Qian}}\ and\ \bibinfo {author} {\bibfnamefont {F.}~\bibnamefont {Liu}},\
  }\href {\doibase 10.1103/physreve.100.062119} {\bibfield  {journal} {\bibinfo
   {journal} {Phys. Rev. E}\ }\textbf {\bibinfo {volume} {100}},\ \bibinfo
  {pages} {062119} (\bibinfo {year} {2019})}\BibitemShut {NoStop}%
\bibitem [{\citenamefont {Brodier}\ \emph {et~al.}(2020)\citenamefont
  {Brodier}, \citenamefont {Mallick},\ and\ \citenamefont
  {de~Almeida}}]{Brodier2020}%
  \BibitemOpen
  \bibfield  {author} {\bibinfo {author} {\bibfnamefont {O.}~\bibnamefont
  {Brodier}}, \bibinfo {author} {\bibfnamefont {K.}~\bibnamefont {Mallick}}, \
  and\ \bibinfo {author} {\bibfnamefont {A.~M.~O.}\ \bibnamefont
  {de~Almeida}},\ }\href {\doibase 10.1088/1751-8121/ab8110} {\bibfield
  {journal} {\bibinfo  {journal} {J. Phys. A: Math. Theor.}\ }\textbf {\bibinfo
  {volume} {53}},\ \bibinfo {pages} {325001} (\bibinfo {year}
  {2020})}\BibitemShut {NoStop}%
\bibitem [{\citenamefont {Qiu}\ \emph {et~al.}(2020{\natexlab{b}})\citenamefont
  {Qiu}, \citenamefont {Fei}, \citenamefont {Pan},\ and\ \citenamefont
  {Quan}}]{Qiu2020a}%
  \BibitemOpen
  \bibfield  {author} {\bibinfo {author} {\bibfnamefont {T.}~\bibnamefont
  {Qiu}}, \bibinfo {author} {\bibfnamefont {Z.}~\bibnamefont {Fei}}, \bibinfo
  {author} {\bibfnamefont {R.}~\bibnamefont {Pan}}, \ and\ \bibinfo {author}
  {\bibfnamefont {H.~T.}\ \bibnamefont {Quan}},\ }\href {\doibase
  10.1103/physreve.101.032113} {\bibfield  {journal} {\bibinfo  {journal}
  {Phys. Rev. E}\ }\textbf {\bibinfo {volume} {101}},\ \bibinfo {pages}
  {032113} (\bibinfo {year} {2020}{\natexlab{b}})}\BibitemShut {NoStop}%
\bibitem [{\citenamefont {Qiu}\ and\ \citenamefont {Quan}(2021)}]{Qiu2021}%
  \BibitemOpen
  \bibfield  {author} {\bibinfo {author} {\bibfnamefont {T.}~\bibnamefont
  {Qiu}}\ and\ \bibinfo {author} {\bibfnamefont {H.-T.}\ \bibnamefont {Quan}},\
  }\href {\doibase 10.1088/1572-9494/ac0813} {\bibfield  {journal} {\bibinfo
  {journal} {Commun. Theor. Phys.}\ }\textbf {\bibinfo {volume} {73}},\
  \bibinfo {pages} {095602} (\bibinfo {year} {2021})}\BibitemShut {NoStop}%
\bibitem [{\citenamefont {Hu}\ \emph {et~al.}(1992)\citenamefont {Hu},
  \citenamefont {Paz},\ and\ \citenamefont {Zhang}}]{Hu1992}%
  \BibitemOpen
  \bibfield  {author} {\bibinfo {author} {\bibfnamefont {B.~L.}\ \bibnamefont
  {Hu}}, \bibinfo {author} {\bibfnamefont {J.~P.}\ \bibnamefont {Paz}}, \ and\
  \bibinfo {author} {\bibfnamefont {Y.}~\bibnamefont {Zhang}},\ }\href
  {\doibase 10.1103/physrevd.45.2843} {\bibfield  {journal} {\bibinfo
  {journal} {Phys. Rev. D}\ }\textbf {\bibinfo {volume} {45}},\ \bibinfo
  {pages} {2843} (\bibinfo {year} {1992})}\BibitemShut {NoStop}%
\bibitem [{\citenamefont {Karrlein}\ and\ \citenamefont
  {Grabert}(1997)}]{Karrlein1997}%
  \BibitemOpen
  \bibfield  {author} {\bibinfo {author} {\bibfnamefont {R.}~\bibnamefont
  {Karrlein}}\ and\ \bibinfo {author} {\bibfnamefont {H.}~\bibnamefont
  {Grabert}},\ }\href {\doibase 10.1103/physreve.55.153} {\bibfield  {journal}
  {\bibinfo  {journal} {Phys. Rev. E}\ }\textbf {\bibinfo {volume} {55}},\
  \bibinfo {pages} {153} (\bibinfo {year} {1997})}\BibitemShut {NoStop}%
\bibitem [{\citenamefont {Ford}\ and\ \citenamefont
  {O'Connell}(2001)}]{Ford2001}%
  \BibitemOpen
  \bibfield  {author} {\bibinfo {author} {\bibfnamefont {G.~W.}\ \bibnamefont
  {Ford}}\ and\ \bibinfo {author} {\bibfnamefont {R.~F.}\ \bibnamefont
  {O'Connell}},\ }\href {\doibase 10.1103/physrevd.64.105020} {\bibfield
  {journal} {\bibinfo  {journal} {Phys. Rev. D}\ }\textbf {\bibinfo {volume}
  {64}},\ \bibinfo {pages} {105020} (\bibinfo {year} {2001})}\BibitemShut
  {NoStop}%
\bibitem [{\citenamefont {Salazar}(2020)}]{Salazar2020}%
  \BibitemOpen
  \bibfield  {author} {\bibinfo {author} {\bibfnamefont {D.~S.~P.}\
  \bibnamefont {Salazar}},\ }\href {\doibase 10.1103/physreve.101.030101}
  {\bibfield  {journal} {\bibinfo  {journal} {Phys. Rev. E}\ }\textbf {\bibinfo
  {volume} {101}},\ \bibinfo {pages} {030101} (\bibinfo {year}
  {2020})}\BibitemShut {NoStop}%
\bibitem [{\citenamefont {Chen}\ \emph {et~al.}(2021)\citenamefont {Chen},
  \citenamefont {Chen}, \citenamefont {Fei},\ and\ \citenamefont
  {Quan}}]{Chen2021a}%
  \BibitemOpen
  \bibfield  {author} {\bibinfo {author} {\bibfnamefont {Y.~H.}\ \bibnamefont
  {Chen}}, \bibinfo {author} {\bibfnamefont {J.-F.}\ \bibnamefont {Chen}},
  \bibinfo {author} {\bibfnamefont {Z.}~\bibnamefont {Fei}}, \ and\ \bibinfo
  {author} {\bibfnamefont {H.~T.}\ \bibnamefont {Quan}},\ }\href@noop {} {\
  (\bibinfo {year} {2021})},\ \Eprint {http://arxiv.org/abs/2108.04128}
  {arXiv:2108.04128 [cond-mat.stat-mech]} \BibitemShut {NoStop}%
\bibitem [{\citenamefont {Ca{\~{n}}izares}\ and\ \citenamefont
  {Sols}(1994)}]{Canizares1994}%
  \BibitemOpen
  \bibfield  {author} {\bibinfo {author} {\bibfnamefont {J.~S.}\ \bibnamefont
  {Ca{\~{n}}izares}}\ and\ \bibinfo {author} {\bibfnamefont {F.}~\bibnamefont
  {Sols}},\ }\href {\doibase 10.1016/0378-4371(94)90146-5} {\bibfield
  {journal} {\bibinfo  {journal} {Phys. A}\ }\textbf {\bibinfo {volume}
  {212}},\ \bibinfo {pages} {181} (\bibinfo {year} {1994})}\BibitemShut
  {NoStop}%
\bibitem [{\citenamefont {Ju}\ \emph {et~al.}(2017)\citenamefont {Ju},
  \citenamefont {Guo},\ and\ \citenamefont {Pan}}]{Ju2017}%
  \BibitemOpen
  \bibfield  {author} {\bibinfo {author} {\bibfnamefont {K.-K.}\ \bibnamefont
  {Ju}}, \bibinfo {author} {\bibfnamefont {C.-X.}\ \bibnamefont {Guo}}, \ and\
  \bibinfo {author} {\bibfnamefont {X.-Y.}\ \bibnamefont {Pan}},\ }\href
  {\doibase 10.1088/0256-307x/34/1/010301} {\bibfield  {journal} {\bibinfo
  {journal} {Chin. Phys. Lett.}\ }\textbf {\bibinfo {volume} {34}},\ \bibinfo
  {pages} {010301} (\bibinfo {year} {2017})}\BibitemShut {NoStop}%
\bibitem [{\citenamefont {Kramers}(1940)}]{Kramers1940}%
  \BibitemOpen
  \bibfield  {author} {\bibinfo {author} {\bibfnamefont {H.~A.}\ \bibnamefont
  {Kramers}},\ }\href {\doibase 10.1016/s0031-8914(40)90098-2} {\bibfield
  {journal} {\bibinfo  {journal} {Physica}\ }\textbf {\bibinfo {volume} {7}},\
  \bibinfo {pages} {284} (\bibinfo {year} {1940})}\BibitemShut {NoStop}%
\bibitem [{\citenamefont {Grabert}\ \emph {et~al.}(1984)\citenamefont
  {Grabert}, \citenamefont {Weiss},\ and\ \citenamefont
  {Talkner}}]{Grabert1984}%
  \BibitemOpen
  \bibfield  {author} {\bibinfo {author} {\bibfnamefont {H.}~\bibnamefont
  {Grabert}}, \bibinfo {author} {\bibfnamefont {U.}~\bibnamefont {Weiss}}, \
  and\ \bibinfo {author} {\bibfnamefont {P.}~\bibnamefont {Talkner}},\ }\href
  {\doibase 10.1007/bf01307505} {\bibfield  {journal} {\bibinfo  {journal} {Z.
  Phys. B}\ }\textbf {\bibinfo {volume} {55}},\ \bibinfo {pages} {87} (\bibinfo
  {year} {1984})}\BibitemShut {NoStop}%
\bibitem [{Note2()}]{Note2}%
  \BibitemOpen
  \bibinfo {note} {Strictly, the substitution requires to amend $\gamma
  _{n},\Lambda _{nm},\xi _{n},\Delta _{nm}$ accordingly, but we only require
  $q_{0}(t)$ and $p_{0}(t)$ to calculate the characteristic function of heat,
  so we skip the further amendment.}\BibitemShut {Stop}%
\end{thebibliography}%

\end{document}